\documentclass[twoside,twocolumn,9pt]{article}
\usepackage{extsizes}
\usepackage[super,sort&compress,comma]{natbib} 
\usepackage[version=3]{mhchem}
\usepackage[left=1.5cm, right=1.5cm, top=1.785cm, bottom=2.0cm]{geometry}
\usepackage{balance}
\usepackage{mathptmx}
\usepackage{sectsty}
\usepackage{graphicx} 
\usepackage{lastpage}
\usepackage[format=plain,justification=justified,singlelinecheck=false,font={stretch=1.125,small,sf},labelfont=bf,labelsep=space]{caption}
\usepackage{float}
\usepackage{fancyhdr}
\usepackage{fnpos}
\usepackage[english]{babel}
\addto{\captionsenglish}{%
	
}
\usepackage{array}
\usepackage{droidsans}
\usepackage{charter}
\usepackage[T1]{fontenc}
\usepackage[usenames,dvipsnames]{xcolor}
\usepackage{setspace}
\usepackage[compact]{titlesec}
\usepackage{hyperref}
\usepackage{amssymb}
\usepackage{chngcntr}

\usepackage{bm}
\usepackage{amsmath}
\usepackage[normalem]{ulem}
\newcommand{\ubar}[1]{\mkern 1.5mu{\uline{{\mkern-1.5mu#1\mkern-1.5mu}}}\mkern 1.5mu}
\newcommand{\ubaro}[1]{\mkern 1.5mu\mbox{\uline{{$\mkern-1.5mu#1\mkern-1.5mu$}}}\mkern 1.5mu}
\newcommand{\uubar}[1]{\ubar{\ubaro{#1}}}
\newcommand{\uuubar}[1]{\ubar{{\ubar{\ubaro{#1}}}}}

\newcommand\uuuline{\bgroup\markoverwith%
	{%
		\textcolor{black}{\rule[-0.5ex]{2pt}{0.4pt}}%
		\llap{\textcolor{black}{\rule[-0.8ex]{2pt}{0.4pt}}}%
		\llap{\textcolor{black}{\rule[-1.15ex]{2pt}{0.4pt}}}%
	}%
	\ULon}

\definecolor{cream}{RGB}{222,217,201}

\newcommand*\diff{\mathop{}\!\mathrm{d}}

\newcommand{\SC}[1]{\textcolor{black} {#1}}

\begin{document}
\pagestyle{fancy}
\thispagestyle{plain}
\fancypagestyle{plain}{
\renewcommand{\headrulewidth}{0pt}
}
\makeFNbottom
\makeatletter
\renewcommand\LARGE{\@setfontsize\LARGE{15pt}{17}}
\renewcommand\Large{\@setfontsize\Large{12pt}{14}}
\renewcommand\large{\@setfontsize\large{10pt}{12}}
\renewcommand\footnotesize{\@setfontsize\footnotesize{7pt}{10}}
\makeatother

\renewcommand{\thefootnote}{\fnsymbol{footnote}}
\renewcommand\footnoterule{\vspace*{1pt}%
	\color{cream}\hrule width 3.5in height 0.4pt \color{black}\vspace*{5pt}} 
\setcounter{secnumdepth}{5}

\makeatletter 
\renewcommand\@biblabel[1]{#1}            
\renewcommand\@makefntext[1]%
{\noindent\makebox[0pt][r]{\@thefnmark\,}#1}
\makeatother 
\renewcommand{\figurename}{\small{Fig.}~}
\sectionfont{\sffamily\Large}
\subsectionfont{\normalsize}
\subsubsectionfont{\bf}
\setstretch{1.125} 
\setlength{\skip\footins}{0.8cm}
\setlength{\footnotesep}{0.25cm}
\setlength{\jot}{10pt}
\titlespacing*{\section}{0pt}{4pt}{4pt}
\titlespacing*{\subsection}{0pt}{15pt}{1pt}

\fancyfoot{}
\fancyfoot[LO,RE]{\vspace{-7.1pt}\includegraphics[height=9pt]{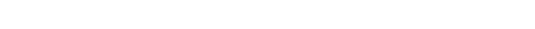}}
\fancyfoot[CO]{\vspace{-7.1pt}\hspace{13.2cm}\includegraphics{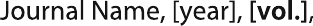}}
\fancyfoot[CE]{\vspace{-7.2pt}\hspace{-14.2cm}\includegraphics{head_foot/RF}}
\fancyfoot[RO]{\footnotesize{\sffamily{1--\pageref{LastPage} ~\textbar  \hspace{2pt}\thepage}}}
\fancyfoot[LE]{\footnotesize{\sffamily{\thepage~\textbar\hspace{3.45cm} 1--\pageref{LastPage}}}}
\fancyhead{}
\renewcommand{\headrulewidth}{0pt} 
\renewcommand{\footrulewidth}{0pt}
\setlength{\arrayrulewidth}{1pt}
\setlength{\columnsep}{6.5mm}
\setlength\bibsep{1pt}

\makeatletter 
\newlength{\figrulesep} 
\setlength{\figrulesep}{0.5\textfloatsep} 

\newcommand{\topfigrule}{\vspace*{-1pt}%
	\noindent{\color{cream}\rule[-\figrulesep]{\columnwidth}{1.5pt}} }

\newcommand{\botfigrule}{\vspace*{-2pt}%
	\noindent{\color{cream}\rule[\figrulesep]{\columnwidth}{1.5pt}} }

\newcommand{\dblfigrule}{\vspace*{-1pt}%
	\noindent{\color{cream}\rule[-\figrulesep]{\textwidth}{1.5pt}} }

\makeatother
\renewcommand{\ULdepth}{1pt}

\twocolumn[
\begin{@twocolumnfalse}
	{\includegraphics[height=30pt]{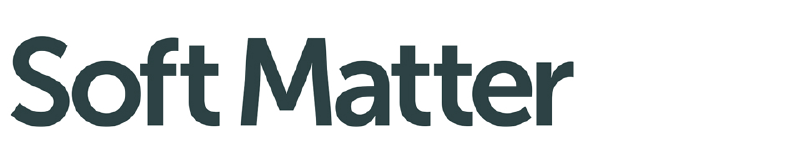}\hfill\raisebox{0pt}[0pt][0pt]{\includegraphics[height=55pt]{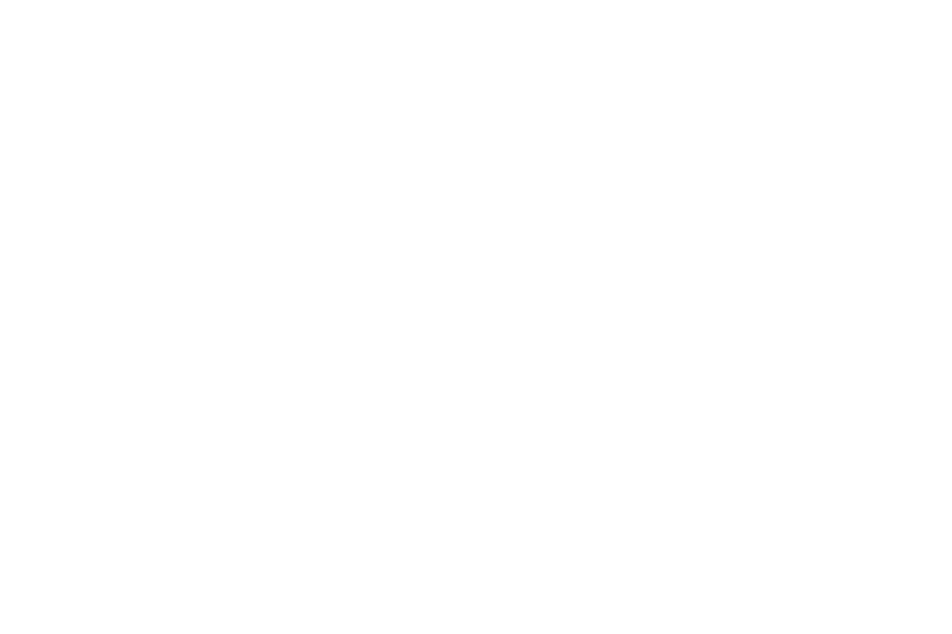}}\\[1ex]
		\includegraphics[width=18.5cm]{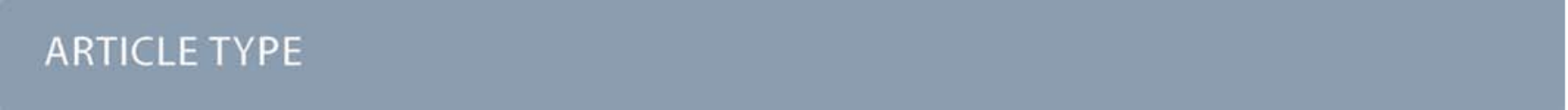}}\par
	\vspace{1em}
	\sffamily
	\begin{tabular}{m{4.5cm} p{13.5cm} }
		
		\includegraphics{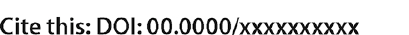} & \noindent\LARGE{\textbf{Effective Medium Theory for Mechanical Phase Transitions of Fiber Networks}} \\
		\vspace{0.3cm} & \vspace{0.3cm} \\
		
		& \noindent\large{Sihan Chen$^{a,b}$, Tomer Markovich$^{b,c,d}$, Fred C. MacKintosh$^{a,b,e,f}$} \\
		
		\includegraphics{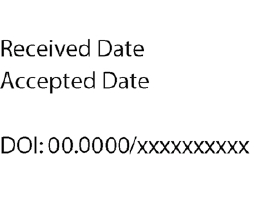} & \noindent\normalsize{Networks of stiff fibers govern the elasticity of biological structures such as the extracellular matrix of collagen. These networks are known to stiffen nonlinearly under shear or extensional strain. Recently, it has been shown that such stiffening is governed by a strain-controlled athermal but critical phase transition, \SC{from a floppy phase below the critical strain to a rigid phase above the critical strain}. While this phase transition has been extensively studied numerically and experimentally, \SC{a complete analytical theory} for this transition remains elusive. Here, we present an effective medium theory (EMT) for this mechanical phase transition of fiber networks. We extend a previous EMT appropriate for linear elasticity to incorporate nonlinear effects via an anharmonic Hamiltonian. The mean-field predictions of this theory, including the critical exponents, scaling relations and non-affine fluctuations qualitatively agree with previous experimental and numerical results.} \\
		
	\end{tabular}
	
\end{@twocolumnfalse} \vspace{0.6cm}

]
\renewcommand*\rmdefault{bch}\normalfont\upshape
\rmfamily
\section*{}
\vspace{-1cm}

\footnotetext{$^a$Department of Physics and Astronomy, Rice University, Houston, TX 77005}
\footnotetext{$^b$Center for Theoretical Biological Physics, Rice University, Houston, TX 77005}
\footnotetext{$^c$School of Mechanical Engineering, Tel Aviv University, Tel Aviv 69978, Israel}
\footnotetext{$^d$Center for Physics and Chemistry of Living Systems, Tel Aviv University, Tel Aviv 69978, Israel}
\footnotetext{$^e$Department of Chemical and Biomolecular Engineering, Rice University, Houston, TX 77005}
\footnotetext{$^f$Department of Chemistry, Rice University, Houston, TX 77005}

\section{Introduction}
Networks formed by stiff fibers are ubiquitous in both biological and artificial materials.~\cite{hough2004viscoelasticity,bryning2007carbon,dan2009continuous,hudnut2018role,van2019emergence,Shivers21037}. One notable example is the extracellular matrix of collagen, which provides structural support to surrounding tissues.~\cite{Sharma2016,hudnut2018role}. These networks possess remarkable mechanical properties, such as stiffening under shear or extensional strain~\cite{fung1967elasticity,Gardel2004,Storm2005191,kabla2007nonlinear,picu2011mechanics,Broedersz2014,vahabi2016elasticity,van2016uncoupling,ban2019strong,burla2019stress,bertula2019strain}, softening under compression~\cite{vahabi2016elasticity,van2016uncoupling} and anomalous Poisson's ratio~\cite{kabla2007nonlinear,vader2009strain,shivers2020nonlinear}. Recently, it has been recognized that the strain stiffening of fiber networks is associated with a critical phase transition: as the applied strain exceeds a critical value, the network transforms from a floppy phase to a rigid phase~\cite{Sharma2016,sharma2016strain,vermeulen2017geometry,jansen2018role,rens2018micromechanical,arzash2019stress,shivers2019scaling,arzash2020finite,arzash2021shear}. Various properties of this mechanical phase transition have been studied through simulations and experiments, including critical exponents~\cite{Sharma2016,sharma2016strain,vermeulen2017geometry,shivers2019scaling,arzash2021shear}, non-affine fluctuations~\cite{shivers2019scaling,arzash2020finite} and scaling relations~\cite{shivers2019scaling,lerner2022scaling}. Despite these efforts, \SC{a complete analytical theory for the nonlinear elasticity  remain elusive.}

Non-affine deformations pose a significant challenge in any analytical description of rheology and mechanical phase transitions. Unlike an affine deformation that corresponds to a uniform deformation field throughout the entire network, non-affine deformations in fiber networks represent inhomogenous  and largely independent deformation of the constituents. This non-affinity has a strong impact on the elasticity of  network, including a significant reduction in the elastic moduli.  Non-affinity also plays a crucial role in mechanical phase transitions. For instance, previous simulations have shown that the non-affine fluctuations diverge at the critical strain.~\cite{shivers2019scaling,arzash2020finite}. As a result, traditional effective medium theory (EMT) cannot be directly applied to fiber networks at a mechanical phase transition since such theories are based on uniform lattices with vanishing non-affinity~\cite{phillips1979topology,thorpe1983continuous,Feng1985,Das2007,broedersz2011criticality,Sheinman2012,Mao2013,Mao20132}. A recent EMT has been proposed to account for non-affine deformations and has shown quantitative agreement with prior linear elasticity of fiber and semiflexible polymer networks~\cite{Linear}. Since it accounts for non-affine deformation, this theory offers a potential framework for describing mechanical phase transitions.

Here, we extend this EMT to describe the nonlinear elasticity of fiber networks. We find good qualitative consistency with prior work on the strain-controlled critical phase transition of fiber networks. To achieve this, we essentially use the EMT of Ref.~\cite{Linear}, with  an extension of an anharmonic effective Hamiltonian. Our theory provides an analytical prediction of the mechanical phase transition, including mean-field critical exponents. Furthermore, it reproduces several qualitative features observed in previous numerical studies, such as the discontinuity of the elasticity~\cite{vermeulen2017geometry,merkel2019minimal,arzash2020finite}, the divergence of non-affine fluctuations~\cite{shivers2019scaling,arzash2020finite} and scaling relations~\cite{shivers2019scaling}. Although our focus is on athermal fiber networks, our theory can also be used as an efficient tool for understanding the influence of thermal fluctuations on such mechanical phase transitions, where numerical simulations are challenging.

\section{Effective Medium Theory}
In this section we construct an EMT for nonlinear elasticity of fiber networks, by extending our previous linear EMT proposed in Ref.~\cite{Linear}. 

\begin{figure}[t]
	\centering
	\includegraphics[scale=0.3]{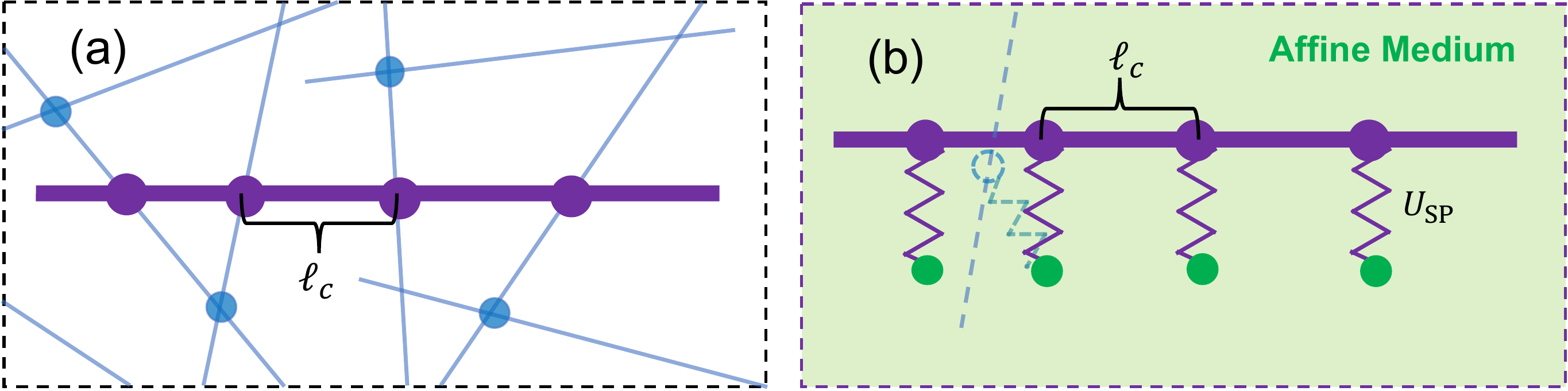}
	\caption{\SC{(a) A 2D sketch of the original  fiber network (in 3D).  Fibers (lines) are connected by  hinge-like crosslinks (dots), with an average crosslinking distance $\ell_c$. The fibers have an average length $L$.    (b) Illustration of the EMT. Each crosslink in the original network is replaced by a spring with potential $U_{\rm SP}$, which connects the fibers with a substrate. The substrate deforms with the same strain as that of the entire network, and is assumed to deform in an affine manner. } }
	\label{fig1}
\end{figure}

Consider a 3D fiber network (the original network) formed by $N$ fibers, each with contour length $L$ (see Fig.~\ref{fig1}(a)). Crosslinks are randomly formed between pairs of fibers with an average distance $\ell_c$. For simplicity the network is assumed to be both isotropic and homogeneous on large scale. The network Hamiltonian can be written as:
\begin{align}
H_O=\sum_{\alpha=1}^N \Big[H_{b}\left[{\bm u}^{\alpha}(s)\right]+ H_{s}\left[\bm u^{\alpha}(s)\right]\Big]\,,\label{e1}
\end{align}
where ${\bm u}^{\alpha}(s)={\bm u}_\parallel^{\alpha}(s)+{\bm u}_\perp^{\alpha}(s)$ is the microscopic displacement of the $\alpha$-th fiber at position $s$ along its contour ($-L/2<s<L/2$), with ${\bm u}_\parallel^{\alpha}(s)$ and  ${\bm u}_\perp^{\alpha}(s)$ being its longitudinal and transverse components, respectively.  $H_{b}\left[{\bm u}(s)\right]=\kappa \int \diff s |\partial ^2{\bm u}_\perp/\partial s^2|^2/2$ and $H_{s}\left[{\bm u}(s)\right]=\mu \int \diff s (|\hat {\bm n}+\partial {\bm u}/\partial s|-1)^2/2$ are the bending and stretching energy, respectively, with $\hat {\bm n}$ being the fiber orientation~\footnote{The form of the stretching energy is different from that in our earlier work~\cite{Linear}. This is because in Ref.~\cite{Linear} we consider linear elasticity only. One can show that for small $\bm u$ the two stretching energies are equivalent.}.  \SC{The form of the bending and the stretching energies are extracted from the classic worm-like chain model of semiflexible polymers~\cite{kratky1949x,aragon1985dynamics,Gittes1998,Morse1998,Broedersz2014}.} If a crosslink exists between the $\alpha$-th and the $\beta$-th fiber, it corresponds to an additional constraint, ${\bm u}^{\alpha}(s_{\alpha\beta})={\bm u}^{\beta}(s_{\beta\alpha})$, with $s_{\alpha\beta}$ ($s_{\beta\alpha}$) being the position of the crosslink on the $\alpha$-th ($\beta$-th) fiber. 

To calculate the elasticity of the original network, in Ref.~\cite{Linear} we have constructed an EMT which reproduces the elasticity of the original network (see Fig.~\ref{fig1}(b)). In this EMT all the fibers in the original networks are conserved, while all crosslinks are replaced by springs that connect the fibers to a substrate, introducing an additional spring energy $H_K$. \SC{The substrate is assumed to deform affinely with the macroscopic deformation tensor $\uubar{\bm \Lambda}_{_{\rm EM}}$ of the network.} The corresponding Hamiltonian is,
\begin{align}
H_{_{\rm EM}}&=\sum_{\alpha=1}^N\Big(H_{b}\left[{\bm v}^{\alpha}(s)\right]+H_{s}\left[\bm v^{\alpha}(s)\right]+H_{K}\left[\bm v_{\rm NA}^{\alpha}(s)\right]\Big)\,,\label{e2}
\end{align}
where the microscopic deformation in the EMT is denoted by ${\bm v}^{\alpha}(s)={\bm v}_A^{\alpha}(s)+{\bm v}_{\rm NA}^{\alpha}(s)$, with ${\bm v}_A^{\alpha}(s)$ being the affine displacement and ${\bm v}_{\rm NA}^{\alpha}(s)$ being the non-affine displacement. The microscopic affine displacements are given by ${\bm v}_A^{\alpha}(s)= s\uubar{\bm \Lambda}_{_{\rm EM}} \cdot\hat{\bm n}^{\alpha}$, with $\hat{\bm n}^{\alpha}$ defining the fiber orientation.  As appropriate for fibers we assume both ${\bm v_{\perp}}$ and ${\bm v_{\parallel}}$ are small such that we can write the stretching energy as: $H_{s}\left[{\bm v}(s)\right]=\mu \int \diff s (\partial \epsilon/\partial s)^2+\hat{\bm n}\cdot(\partial {\bm v}_\parallel/\partial s)|\partial {\bm v}_\perp/\partial s|^2/4$,  where $\partial \epsilon/\partial s=\hat{\bm n}\cdot(\partial {\bm v}_\parallel/\partial s)+|\partial {\bm v}_\perp/\partial s|^2/2$. Then, to simplify the calculation we neglect the coupling between the transverse and longitudinal displacements and approximate the energy as: $H_{s}\left[{\bm v}(s)\right]=\mu \int \diff s (\partial \epsilon/\partial s)^2$. For simple shear deformation, neglecting this coupling term can be interpreted as a mean-field approximation: since half of the filaments are stretched while the other half are compressed, taking the mean-field value of $\hat{\bm n}\cdot(\partial {\bm v}_\parallel/\partial s)$ (which is zero) results in a vanishing coupling term between the longitudinal and transverse displacements. \SC{$H_K$ is a functional of the non-affine displacements, because only non-affine displacements can introduce relative displacements between the fibers and the affine substrate, which stretch the springs.}  $H_K$ can also be written as a summation of spring energy at all crosslinking positions,
\begin{align}
H_{K}\left[\bm v_{\rm NA}^{\alpha}(s)\right] = \sum_i U_{\rm SP}[ v_{\rm NA}^{\alpha}(s^\alpha_i)]\,,
\label{e3}
\end{align}
with $U_{\rm SP}$ being the energy of a single spring and $s^\alpha_i$ is the position of the $i$-th crosslink on the $\alpha$-th fiber. 

We are interested in how the two networks (original and EMT) respond to \SC{macroscopic} deformations $\uubar{\bm \Lambda}_{_O}$ and $\uubar{\bm \Lambda}_{_{\rm EM}}$, respectively.  In this work we focus on simple shear deformation that is most commonly studied in rheology of fiber networks. Other deformation types, including uniaxial and bulk strain can lead to mechanical phase transition as well~\cite{Sheinman2012,shivers2020nonlinear,arzash2022mechanics,lee2023partition}.  Without loss of generality, we assume simple shear in the $x-z$ plane in the $x$ direction, with $\gamma_{_O}$ and $\gamma_{_{\rm EM}}$ being the nonzero $xz$ component of each of the \SC{macroscopic} deformation tensors \SC{and other components constrained to zero.}  The deformations result in \SC{macroscopic} stresses in the two networks, with $\sigma_{_O}$ and $\sigma_{_{\rm EM}}$ being the $xz$ (and $zx$) component of each of the stress tensors~\footnote{In principle the stress tensor may have other non-zero components, e.g., normal stresses~\cite{janmey2007negative,heussinger2007nonaffine,conti2009cross,shivers2019normal}.}. \SC{In such a setup, and because fiber networks are elastic and athermal, there is a one-to-one relation between $\sigma$  and $\gamma$ in each network. To identify this relation, one can either calculate $\sigma$ at a given $\gamma$ (strain-controlled) or calculate $\gamma$ at a given $\sigma$ (stress-controlled). The two methods are equivalent and are related by a Legendre transform.  Here we choose the stress-controlled scenario, in which the network deformation at a given shear stress} is found from the minimum-energy states of the two total energies, $E_{_O}=H_{_O}-V\sigma_{_O}\gamma_{_O}$ and $E_{_{\rm EM}}=H_{_{\rm EM}}-V\sigma_{_{\rm EM}}\gamma_{_{\rm EM}}$. $V$ is the system volume that is set to unity from here on. As discussed above, an additional constraint is that, except for the $xz$ components, all other components of the deformation tensors must be zero.  \SC{Importantly, while the two strains $\gamma_{_O}$ and $\gamma_{_{\rm EM}}$ describe the macroscopic strain of the corresponding networks (and also the substrate of the EMT), each  fiber has a different, non-affine microscopic deformation $\bm u^\alpha$ and $\bm v^\alpha$. The macroscopic and microscopic deformations are intrinsically related to each other such that $\gamma_{_O}=\gamma_{_O}(\{\bm u^\alpha\})$ and $\gamma_{_{\rm EM}}=\gamma_{_{\rm EM}}(\{\bm v^\alpha\})$, see Eq.~(\ref{e7}) below for details. Therefore, the total energies $E_O$ and $E_{_{\rm EM}}$ are also functionals of microscopic variables $\bm u^\alpha$ and $\bm v^\alpha$.} The values of $\bm u^\alpha$ and $\bm v^\alpha$ in the minimum energy state are $\widetilde{{\bm u}}^{\alpha}$ and $\widetilde{{\bm v}}^\alpha$, for the original network and the EMT network, respectively. 

The nonlinear elasticity of two networks is quantified using the differential shear moduli $K_{_{O}}= \partial \sigma_{_{O}}/\partial \gamma_{_{O}}$ and $K_{_{\rm EM}}= \partial \sigma_{_{\rm EM}}/\partial \gamma_{_{\rm EM}}$. Our goal is to construct an EMT network that reproduces the differential shear modulus of the original network, {i.e., } $K_{_{\rm EM}}=K_{_{O}}$. For this we rewrite $1/K_{_{\rm EM}}=1/K_{_{O}}$ using the chain rule, 
	\begin{equation}
\sum_{\alpha i}\frac{\partial \widetilde {\bm u}^{\alpha}_{i}}{\partial \sigma_{_O}}\cdot\frac{\partial \gamma_{_O}}{\partial \widetilde {\bm u}^{\alpha}_{i}}=\sum_{\alpha i}\frac{\partial \widetilde {\bm v}^{\alpha}_{i}}{\partial \sigma_{_{\rm EM}}}\cdot \frac{\partial \gamma_{_{\rm EM}}}{\partial \widetilde {\bm v}^{\alpha}_{i}}\,,\label{e6}
	\end{equation}
with $\widetilde{{\bm u}}^{\alpha}_i$ and $\widetilde{{\bm v}}^\alpha_i$ being the displacements of the $i$-th crosslink of the $\alpha$-th fiber in the original network and the EMT network, respectively. \SC{Equation~(\ref{e6}) decomposes the network mechanical response into two parts: When a macroscopic stress ($\sigma_{_O}$ or $\sigma_{_{\rm EM}}$) is imposed, it leads to microscopic deformations of each fiber ($\widetilde{{\bm u}}$ or $\widetilde{{\bm v}}$); These microscopic deformations further determine the macroscopic deformation of the entire network ($\gamma_{_O}$ or $\gamma_{_{\rm EM}}$).} Due to the random crosslinks in the original network, let 
\begin{subequations}\label{e8}
	\begin{equation}\label{e8a}
	\frac{\partial \widetilde {\bm u}^{\alpha}_{i}}{\partial \sigma_{_O}}=\left\langle\frac{\partial \widetilde {\bm u}^{\alpha}_{i}}{\partial \sigma_{_O}}\right\rangle+\bm\xi^\alpha_i\,,
	\end{equation}
	\begin{equation}\label{e8b}
	\frac{\partial \gamma_{_O}}{\partial \widetilde {\bm u}^{\alpha}_{i}}=\left\langle\frac{\partial \gamma_{_O}}{\partial \widetilde {\bm u}^{\alpha}_{i}}\right\rangle+\bm\eta_i^\alpha\,, 
	\end{equation}
\end{subequations} 
where $\langle...\rangle$ denotes averages with respect to random crosslinking angles. $\xi^\alpha_i$ and $\eta^\alpha_i$ are two noise-like terms describing the effects of random crosslinks, with $\langle \xi^\alpha_i\rangle=\langle \eta^\alpha_i\rangle=0$. In the thermodynamic limit, Eq.~(\ref{e6}) is rewritten as
	\begin{equation}
\sum_{\alpha i}\left\langle\frac{\partial \widetilde {\bm u}^{\alpha}_{i}}{\partial \sigma_{_O}}\right\rangle\cdot\left\langle\frac{\partial \gamma_{_O}}{\partial \widetilde {\bm u}^{\alpha}_{i}}\right\rangle+\sum_{\alpha i}\langle \bm\xi^\alpha_i\cdot \bm\eta^\alpha_i\rangle=\sum_{\alpha i}\frac{\partial \widetilde {\bm v}^{\alpha}_{i}}{\partial \sigma_{_{\rm EM}}}\cdot \frac{\partial \gamma_{_{\rm EM}}}{\partial \widetilde {\bm v}^{\alpha}_{i}}\,.\label{e11}
\end{equation}
Here we ignore the correlation between two noise-like terms by assuming $\langle \bm\xi^\alpha_i\cdot \bm\eta^\alpha_i\rangle=0$. To ensure that Eq.~(\ref{e11}) holds for any strain and stress, we let our EMT satisfy the following criterion
\begin{subequations}\label{e5}
	\begin{equation}\label{e5a}
	\left\langle\frac{\partial \widetilde {\bm u}^{\alpha}_{i}}{\partial \sigma_{_O}}\right\rangle=\frac{\partial \widetilde {\bm v}^{\alpha}_{i}}{\partial \sigma_{_{\rm EM}}}\,,
	\end{equation}
	\begin{equation}\label{e5b}
	\left\langle\frac{\partial \gamma_{_O}}{\partial \widetilde {\bm u}^{\alpha}_{i}}\right\rangle=\frac{\partial \gamma_{_{\rm EM}}}{\partial \widetilde {\bm v}^{\alpha}_{i}}\,. 
	\end{equation}
\end{subequations} 
As noted in Ref.~\cite{Linear}, such requirements may be stronger than needed. However, assuming solutions to these combined equations can be found, these must agree with Eq.~(\ref{e6}). Thus, by construction, if one finds an EMT that obeys Eq.~(\ref{e5}), it is guaranteed that Eq.\ \eqref{e6} holds~\footnote{There is, in general, the possibility that one will not be able to find solution to Eq.~(\ref{e5}), while a solution to Eq.~(\ref{e6}) does exist. As we show below, we find a solution to Eq.~(\ref{e5}), thus Eq.~(\ref{e6}) is obeyed as required.}.  Equation (\ref{e5a}), in particular, is natural in that it is equivalent to the coherent potential approximation (CPA) used in prior EMTs~\cite{Feng1985}. 

To approximate the differential elasticity for arbitrary strain and stress, Eqs.~(\ref{e5a}, \ref{e5b}) should hold when the derivatives are evaluated at any deformation. In our linear EMT of Ref.~\cite{Linear} the partial differentials are evaluated at the undeformed state, which gives the spring energy in the linear regime~\cite{Linear},
\begin{align}
U_{\rm SP}(\bm v_{\rm NA})=\frac{9\kappa}{\ell_c^3}|\bm v_{\rm NA}|^2. 
\label{e4}
\end{align}
In such a linear regime, Eq.~(\ref{e5b}) leads to a relation between the macroscopic and microscopic deformations, 
\begin{equation}
\begin{aligned}
\uubar{\bm \Lambda}_{_{\rm EM}}=\sum_\alpha \int_{-L/2}^{L/2} \diff s\,{{\bm v}^\alpha(s)\cdot{\uuubar {\bm T}}(\hat{\bm n}^\alpha,s) },
\end{aligned}
\label{e7}
\end{equation} 
where 
 \begin{equation}
\begin{aligned}
\uuuline {\bm T}(\hat {\bm n},s) = [f_\parallel(s)-f_\perp(s)]\hat {\bm n}\hat {\bm n}\hat {\bm n} +f_\perp(s) \uubar{\bm I} \hat {\bm n}\,,
\end{aligned}
\label{S209}
\end{equation} 
with 
 \begin{equation}
\begin{aligned}
&f_{\parallel}(s) =  \frac{3}{NL}[\delta(s-L/2)-\delta(s+L/2)]\,,\\
&f_{\perp}(s) = \frac{36}{NL^3}s\,.
\end{aligned}
\label{S210}
\end{equation} 
Here $\delta(s) $ is the Dirac-$\delta$ function~\footnote{\SC{Note that the integral in Eq.~(\ref{e7}) is performed from $s=(-L/2)^-$ to $s=(L/2)^+$, such that the two $\delta$ functions are covered in the integral. }}.
The value of $\gamma_{_{\rm EM}}$  can be found by minimizing the total energy with respect to ${\bm v}^{\alpha}$ ($\gamma_{_{\rm EM}}$ is determined from ${\bm v}^{\alpha}$ according to Eq.~(\ref{e7})), 
\begin{equation}
\begin{aligned}
E_{_{\rm EM}} = H_{_{\rm EM}}-\gamma_{_{\rm EM}} \sigma_{_{\rm EM}}\,.
\end{aligned}
\label{e9}
\end{equation} 
\SC{In this work we assume isotropic networks and the filament orientations $\hat{\bm n}^\alpha$ is sampled from an isotropic distribution. In Appendix~\ref{Ac} we detail the energy minimization process.}

In Ref.~\cite{Linear} we show that the linear version of the EMT successfully predicts the linear shear modulus $G$ of fiber networks. Here, we study the nonlinear elasticity and consider the nonlinear version of Eqs.~(\ref{e5a},\ref{e5b}), in which all derivatives may depend on the network deformation. We assume $\gamma\ll1$ such that the relation between $\gamma$ and $\bm u,\bm v$ (Eq.~(\ref{e5b})) is still linear. This assumption is appropriate because it is known that fiber networks can exhibit nonlinear elasticity at small strain~\cite{Broedersz2014}, as we discuss in detail below in Sec.~\ref{sec5}. Therefore, we neglect non-linearity in Eq.~(\ref{e5b}) and only treat non-linearity in Eq.~(\ref{e5a}). This results in an anharmonic correction term $U_{\rm AH}$ to the spring energy:
\begin{align}
U_{\rm SP}(\bm v_{\rm NA})=\frac{9\kappa}{\ell_c^3}|\bm v_{\rm NA}|^2+U_{\rm AH}(\bm v_{\rm NA}),
\label{e10} 
\end{align}
which consequently  gives rise to nonlinear elasticity of the network. As we show below in Sec.~\ref{sec3}, $U_{\rm AH}$ is independent of $\kappa$ in the small $\kappa$ limit. This is consistent with previous numerical results in which the nonlinear regime is dominated by the stretching energy~\cite{Sharma2016}. In Sec.~\ref{sec3} we derive the form of $U_{\rm AH}$. In Sec.~\ref{sec4} we show that the anharmonic spring energy leads to the mechanical phase transition, and then predict the mean-field critical exponents and non-affine fluctuations. 

\section{Elastic energy of nonlinear springs}
\label{sec3}
We continue by calculating $U_{\rm AH}$ using Eq.~(\ref{e5a}). As we have pointed out in Ref.~\cite{Linear}, the longitudinal part in the spring energy only slightly affects the network deformation in the linear regime, because the longitudinal displacements of fibers are always restricted by the stretching energy, even in the absence of the springs. In the nonlinear regime, the longitudinal part of the spring energy is also not important: the nonlinear stiffening of the network corresponds to a transition from a bend-dominated regime to a stretch-dominated regime. Because of the large stretching energy of each fiber, the longitudinal displacements of the fibers are always stretch-dominated and have little contribution to the stiffening of the network. Therefore, we neglect the energy in $U_{\rm AH}$ due to the longitudinal displacement and write $U_{\rm AH}$ as $U_{\rm AH}({\bm v}_{\rm NA\perp})$. 
\subsection{Infinite molecular weight limit: $L/\ell_c\to\infty$}
To analyze Eq.~(\ref{e5a}), we adopt the same method as we used in Ref.~\cite{Linear}, the coherent potential approximation: Because the stress can be decomposed into tensions on each node, we exert a test force $\bm F$ on a particular crosslink on the same fiber of both the original network and the EMT, and measure the resulting displacements $\delta {\bm r}_O$ and $\delta{\bm r}_{\rm EM}$, respectively.  By letting $\langle \delta {\bm r}_O\rangle_{\hat{\rm n}}=\delta{\bm r}_{\rm EM}$, where $\hat{\bm n}$ is the orientation of the other fiber connected to the crosslink in the original network, one obtain the form of $U_{\rm AH}$. Note that in Ref.~\cite{Linear} we assumed small $\bm F$ because we were only interested in the linear elasticity. Here we extend the procedure to any finite $\bm F$. 

In the infinite molecular weight limit ($L/\ell_c\to \infty$), the crosslinks adjacent to the particular crosslink can be treated as being fixed, because in this limit the network is densely crosslinked, making it energetically unfavorable to move the adjacent crosslinks.  The resulting $U_{\rm \rm AH}$ is (see Appendix~\ref{A1} for detailed derivation):
\begin{align}
U_{\rm AH}(\bm v_{\rm NA\perp})=2.32\frac{\mu}{\ell_c^3}|\bm v_{\rm NA\perp}|^4. 
\label{e301}
\end{align}
The quartic term in Eq. (\ref{e301}) suggests that the energy has a stronger-than-harmonic dependence on the displacement, which is consistent with the expectation from Ref.~\cite{lerner2022scaling}. This means that the spring stiffens non-linearly as the displacement increases. The coefficient in Eq.~(\ref{e301}) is insensitive to $\kappa$ because in the derivation we only keep the leading order term in the elastic energy: The bending energy contributes  an additional anharmonic term $\sim(\kappa/\ell_c^5)|\bm v_{\rm NA\perp}|^4$, but it is negligible because $\kappa\ll\mu\ell_c^2$, i.e., for bend-dominated compliance for which the transition is apparent\cite{Sharma2016}. Note that Eq.~(\ref{e301}) is only valid in the high molecular weight limit. In the finite molecular weight limit, the spring energy needs to be corrected, as detailed below. 

\subsection{Finite molecular weight: $L/\ell_c<\infty$}
We now derive $U_{\rm AH}$ for finite molecular weight. Let us start with a simple limit of central-force networks ($\kappa=0$).
For central-force networks, an important observation in previous studies is the emergence of the strain-controlled phase transition: The stiffening of the network only happens for strain above a critical value $\gamma_c$, and the network has zero elasticity for strain below $\gamma_c$. This suggests that for central-force networks the network free energy is singular at $\gamma_c$, which also implies a singular $U_{\rm AH}$:
\begin{equation}
\label{e307}
U_{\rm AH}(\bm v_{\rm NA\perp})=\left\{
\begin{aligned}
&0  & |\bm v_{\rm NA\perp}|\leq v_c \\
&2.32\frac{\mu}{\ell_c^3}|\bm v_{\rm NA\perp}|^4-\Delta U_{\rm AH}(\bm v_{\rm NA\perp})   & |\bm v_{\rm NA\perp}|>v_c.
\end{aligned}
\right.
\end{equation}
Here $U_{\rm AH}$ is singular at a loop $|\bm v_{\rm NA\perp}|=v_c$ because of the network rotational symmetry. $\Delta U_{\rm AH}$ is a finite-molecular-weight correction of $U_{\rm AH}$ which vanishes when $L/\ell_c\to \infty$.  When deriving the spring energy in the infinite molecular weight limit, we assume that the crosslinks adjacent to the deformed crosslink are fixed. Such an assumption is inappropriate for finite molecular weight, due to the reduced number of constraints imposed by crosslinks. In this finite molecular weight case, the displacements of the adjacent crosslinks effectively reduce the spring energy compared to Eq.~(\ref{e301}). The leading term of $\Delta U_{\rm AH}$ is
\begin{equation}
\begin{aligned}
\Delta U_{\rm AH}=c\frac{\mu}{\ell_c}|\bm v_{\rm NA\perp}|^2, 
\end{aligned}
\label{e302}
\end{equation} 
which is quadratic due to the network isotropy. The coefficient $c$ is a dimensionless number. 

\begin{figure}[t]
	\centering
	\includegraphics[scale=0.34]{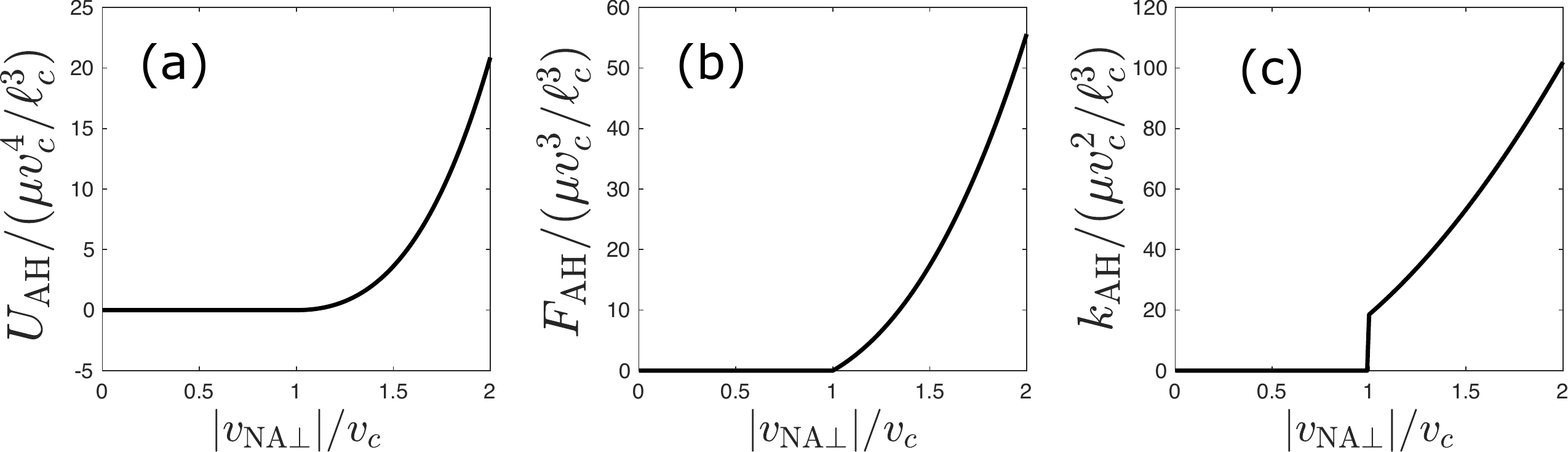}
	\caption{(a) Anharmonic spring energy $U_{\rm AH}$ calculated from  Eq.~(\ref{e304}). (b) Force-extension relation of the anharmonic spring calculated from Eq.~(\ref{e303}). (c) Differential spring constant of the anharmonic spring calculated from Eq.~(\ref{e305}).   }
	\label{spring}
\end{figure}

Remarkably, the second part of Eq.~(\ref{e307}) has the form of a meanfield free energy in Landau theory for second-order phase transition (see Fig.~\ref{spring}(a)). For this potential we can define a force-extension relation $\bm F_{\rm AH}=\diff U_{\rm AH}/\diff \bm v_{\rm NA\perp}$, which must be continuous, as a discontinuity in the network stress is unphysical. This statement can be proved by contradiction: Assume the stress is discontinuous at $\gamma_c$, i.e., let the stress at $\gamma = \gamma_c^-$ be $0$ and the stress at $\gamma = \gamma_c^+$ be $\sigma_c$. Imagine a network starting with strain $\gamma_c^-$ and zero stress, where the stress then quasistatically increases from $0$ to $\sigma_c$. During this process the network strain remains unchanged, such that no external work is done. After the process, because of the finite stress, some of the fibers in the network must be stretched and the network gains a non-zero elastic energy. This process contradicts the first law of thermodynamics, because the internal energy of the system increases with no external work done (there is no heat transfer because the temperature is zero). Therefore, both the stress and $\bm F_{\rm AH}$ must be continuous at the critical point\footnote{This statement may not be true for non-elastic systems, e.g., plastic systems, for which energy can be dissipated. In such systems a discontinuous yield stress may emerge~\cite{chakrabarty2006}. }, leading to $c = 4.64 v_c^2/\ell_c^2$ and 
\begin{equation}
\begin{aligned}
U_{\rm AH}(\bm v_{\rm NA\perp}) &= 2.32 \frac{\mu}{\ell_c^3}(|\bm v_{\rm NA\perp}|^2-v_c^2)^2\Theta(|\bm v_{\rm NA\perp}|^2-v_c^2)\,,
\end{aligned}
\label{e304}
\end{equation} 
where $\Theta(x)$ is the Heavyside function. The corresponding $\bm F_{\rm AH}$ is
\begin{equation}
\begin{aligned}
\!&\bm F_{\rm AH}(\bm v_{\rm NA\perp}) = 9.27\frac{\mu}{\ell_c^3}\bm v_{\rm NA\perp}(|\bm v_{\rm NA\perp}|^2-v_c^2)\Theta(|\bm v_{\rm NA\perp}|^2-v_c^2)\,.
\end{aligned}
\label{e303}
\end{equation} 
Equation (\ref{e303}) suggests that, even an infinitesimal force leads to a displacement with magnitude $v_c$ (see Fig.~\ref{spring} (b)). As we show below in Sec.~\ref{sec4}, at the network level it corresponds to the mechanical phase transition, in which an infinitesimal stress leads to a strain with magnitude $\gamma_c\sim v_c/L$.

We then define a differential spring constant, $k_{\rm AH}=  \diff^2U_{\rm AH}/\diff |\bm v_{\rm NA\perp}|^2$, which according to  Eq.~(\ref{e304}) gives
\begin{equation}
\begin{aligned}
k_{\rm AH}(\bm v_{\rm NA\perp}) &= 9.27\frac{\mu}{\ell_c^3}(3|\bm v_{\rm NA\perp}|^2-v_c^2)\Theta(|\bm v_{\rm NA\perp}|^2-v_c^2)\,.
\end{aligned}
\label{e305}
\end{equation} 
There is a discontinuity in $k_{\rm AH}$ at $|\bm v_{\rm NA\perp}|=v_c$, where $k_{\rm AH}$ jumps from $0$ to $k_c = 18.54\mu v_c^2/\ell_c^3$ (see Fig.~2 (c)). Although the stress must be continuous, the stiffness needs not be. Such a discontinuity in stiffness can be understood as the first appearance of a state of self stress \cite{vermeulen2017geometry}.

The range over which the anharmonic spring is floppy to displacement is characterized by $v_c$. To estimate $v_c$, we adopt the self-consistent approach proposed in Ref.~\cite{Licup2016}: Consider the original network, in which each crosslink on a particular filament can have a displacement $v_c$ without costing any energy, i.e., each segment is free to rotate to an angle $\theta\sim v_c/\ell_c$. The direction of such rotation is random due to the random crosslinking angles. For each segment, the rotation leads to a reduction of the projected length on the fiber backbone, $\Delta \ell \sim \ell_c \theta^2\sim v_c^2/\ell_c$. The change in the end-to-end distance of the fiber is the sum of the change of the projected length of $L/\ell_c$ segments, $\Delta L\sim(L/\ell_c)(v_c^2/\ell_c)$. The self-consistent criterion imposes $v_c=\Delta L$, leading to
\begin{equation}
\begin{aligned}
v_c\sim\ell_c^2/L\,\, \,\,\,\,{\rm or}\,\,\,\,\,\,
\gamma_c\sim v_c/L\sim \ell_c^2/L^2.
\end{aligned}
\label{e306}
\end{equation} 
Equation~(\ref{e306}) suggests that when  $L/\ell_c\to\infty$, $v_c\to0$ and Eq.~(\ref{e304}) reduces to Eq.~(\ref{e301}). This is consistent with our assumption that in the infinite molecular weight limit the displacement of adjacent crosslinks can be neglected. 

Above we have derived $U_{\rm AH}$ for central-force networks ($\kappa=0$). In principle,  for finite $\kappa$ there should be a correction of $U_{\rm AH}$ due to the bending energy. Because the bending and stretching energies are additive, for $\kappa\ll\mu\ell_c^2$ we expect such a correction term to be unimportant to $|\bm v_{\rm NA\perp}|$ both below and above $v_c$. For $|\bm v_{\rm NA\perp}|<v_c$, the leading correction term is a quartic term $~(\kappa/\ell_c^5) |\bm v_{\rm NA\perp}|^4$, because all quadratic terms are accounted for in the calculation of the linear elasticity. This term is much smaller than the quadratic term $\sim(\kappa/\ell_c^3) |\bm v_{\rm NA\perp}|^2$ in $U_{\rm SP}$ because we assume $\gamma\ll 1$ such that $|\bm v_{\rm NA\perp}|\ll \ell_c$. For  $|\bm v_{\rm NA\perp}|\geq v_c$, the leading correction term is a quadratic term $(\kappa/\ell_c^3) |\bm v_{\rm NA\perp}|^2$, which is much smaller than $|\Delta U_{\rm AH}|$ since $\kappa\ll\mu\ell_c^2$. Therefore, in any case the correction of $U_{\rm AH}$ due to bending energy can be neglected, and we use Eq.~(\ref{e304}) as the anharmonic part of the spring energy for any $\kappa$. 

\section{Results}
\label{sec4}
\subsection{Network Nonlinear Elasticity }
Having identified  $U_{\rm AH}$, we substitute Eq.~(\ref{e304}) into Eq.~(\ref{e10}) to obtain the complete  spring energy $U_{\rm SP}$. The EMT network strain $\gamma_{_{\rm EM}}$ under a given stress $\sigma_{_{\rm EM}}$ is found by numerically minimizing Eq.~(\ref{e9}) with respect to $\bm v^\alpha$, see Appendix~\ref{Ac} for details. The network nonlinear elasticity is then calculated with $K_{_{\rm EM}}=\diff \sigma_{_{\rm EM}}/\diff \gamma_{_{\rm EM}}$. In Fig.~3 (a) we plot $K_{_{\rm EM}}$ as function of $\gamma_{_{\rm EM}}$ for various $\kappa$ values. We start with the $\kappa=0$ case. In this non-bending limit, the network has a vanishing shear modulus for $\gamma_{_{\rm EM}}<\gamma_c$. As expected from our construction, when the network reaches the critical point $\gamma=\gamma_c$, it immediately gains a non-zero elasticity $K_{_{\rm EM}}=K_c$, i.e., $K_{_{\rm EM}}$ is discontinuous at the critical point. Such a discontinuity is a result of the discontinuous differential spring constant, see Eq.~(\ref{e305}) and Fig.~\ref{spring}(c). This is consistent with the mechanical phase transition being second-order because $K_{_{\rm EM}}$ is a second derivative of the free energy, similar to the heat capacity in a temperature-controlled transition. This is also consistent with a recent scaling theory and prior numerical results~\cite{shivers2019scaling,vermeulen2017geometry,arzash2020finite}.  

\begin{figure}[t]
	\centering
	\includegraphics[scale=0.4]{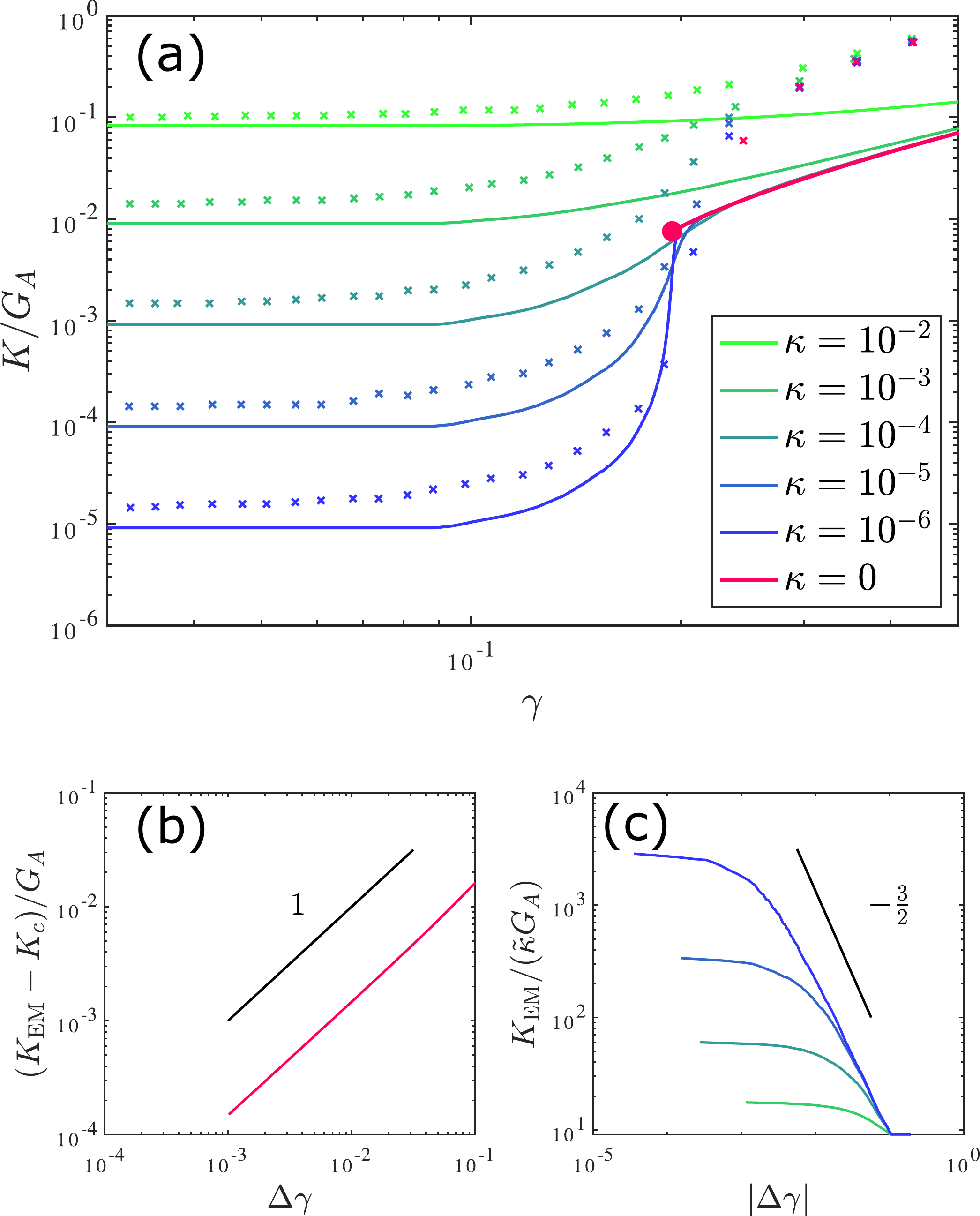}
	\caption{\SC{(a) Differential elastic modulus as function of strain $\gamma$, for different rescaled bending rigidity, $\tilde \kappa = \kappa /(\mu \ell_c^2)$. Lines are theoretical predictions of the EMT with $L=5\ell_c$ and $v_c = 0.05 \ell_c$. The red dot denotes a discontinuous transition at the critical point, $(\gamma_c,K_c)$, for $\tilde \kappa = 0$. $G_A = NL\mu/15$ is the affine modulus. Crosses are simulation results of 3D diluted phantom fcc lattice with connectivity $Z=3.2$ (equivalent to $L=5\ell_c$), reproduced from Ref.~\cite{Licup2016}. No fitting parameter is present (the value of $v_c$ in the EMT is chosen to ensure the same critical strain $\gamma_c\approx 0.2$ as in the simulation.) } (b) Scaling behavior of the EMT elasticity for $\gamma>\gamma_c$, with $\Delta \gamma = \gamma-\gamma_c$ and $\tilde \kappa=0$. (c) Scaling behavior for $\gamma<\gamma_c$ and for various $\tilde \kappa$ values from $10^{-6}$ to $10^{-3}$.  The color coding and EMT parameters used in (b) and (c) is the same as that in (a). }
	\label{K}
\end{figure}

We now consider finite $\kappa$ values, which stabilize the network and introduce non-zero linear elasticity $G_{\rm EM}\sim\kappa$. Below the critical point ($\gamma<\gamma_c$) is the bend-dominated regime, in which $K_{_{\rm EM}}\sim\kappa$. The network still undergoes a stiffening near $\gamma=\gamma_c$, although this is now smooth with no phase transition. The stiffening, however, becomes increasingly rapid as $\kappa$ is reduced. Above the critical point ($\gamma>\gamma_c$), the modulus becomes independent of $\kappa$ for  $\kappa\ll\mu\ell_c^2$, consistent with a stretch-dominated regime. These qualitative features agree with previous simulations of both 2D and 3D networks~\cite{Sharma2016,sharma2016strain,vermeulen2017geometry,arzash2021shear}. \SC{For comparison, previous simulation results of 3D phantom fcc lattice are also plotted in Fig.~\ref{K}(a). The EMT qualitatively captures the linear and the nonlinear elasticity of the simulations. However, we find that it underestimates the stiffening. This is not surprising because the  stiffening in real networks is expected to be governed by non mean-field behavior, which cannot be reproduced by our meanfield EMT, see Sec.~\ref{sec4.2} below. Therefore, we only expect qualitative agreement between the EMT and simulations. While the deviation is small both near the critical point and in the floppy phase, it becomes considerable when the the strain is far above $\gamma_c$. This is possibly due to the geometric nonlinearity that is neglected in the EMT. For large strain ($\gamma\approx 1$) the effect of the geometric nonlinearity can be strong. }\SC{Interestingly, the elasticity at the critical point $K(\gamma_c)$ was recently studied using numerical simulation~\cite{lerner2022scaling}, which reported that $K(\gamma_c)$ in the small-$\kappa$ limit is significantly smaller than $K(\gamma_c)$ for zero $\kappa$, indicating  an underlying difference between the two cases. This difference is also reproduced by our theory, see Fig.~\ref{Fig.A3} in Appendix \ref{Ad}.}

While the signal of the phase transition is apparent for $\kappa=0$ because of the discontinuity in $K_{_{\rm EM}}$, for any finite $\kappa$ value, $K_{_{\rm EM}}$ becomes continuous and the criticality is less obvious. For large $\kappa$ values the phase transition is unidentifiable, see Fig.~\ref{K}(a). Since biopolymer networks always have finite $\kappa$ values, it is important to identify the range of $\kappa$ in which the criticality dominates the nonlinear stiffening. For the criticality to be identifiable, we need $G_{\rm EM}\ll K_c$. For each individual spring this suggests that $\kappa/\ell_c^3\ll k_c$ (see Eq.~(\ref{e305})), which further leads to $\kappa\ll\mu\ell_c^4/L^2$. For $\kappa\gtrsim\mu\ell_c^4/L^2$, the criticality becomes unimportant and the analytic form of $U_{\rm AH} $ in Eq.~(\ref{e301}) is sufficient in describing the stiffening.  \SC{For real fibers with diameter $2a$, we expect $\kappa\sim a^4$ and $\mu \sim a^2$, and the criterion $\kappa\ll\mu\ell_c^4/L^2$ reduces to $a\ll \ell_c^2/L$. This can be easily satisfied, given that $a$ is  of order 10-100 $\rm nm$ and both $\ell_c $ and $L$ are of order micrometers, e.g., in collagen networks.  }

\subsection{Critical Exponents}
\label{sec4.2}
The critical behavior of fiber networks have been studied in both simulation and experiments. Examples of such behavior for the differential modulus include $K-K_c\sim\mu|\Delta \gamma|^f$ for $\Delta \gamma = \gamma-\gamma_c>0$ and $K\sim\kappa |\Delta \gamma|^{f-\phi}$~\cite{Sharma2016,Rens2018,shivers2019scaling,merkel2019minimal,arzash2021shear}. 
Simulations have identified various values of these exponents, with $0<f<1$ and $\lambda=\phi-f\sim1.5$. 
Our EMT predicts mean-field exponents, $f= 1$ and $\lambda=1.5$ (see Fig.~\ref{K} (c) and (d)), which agree with Refs.\ \cite{Rens2018,merkel2019minimal}. On the other hand, extensive simulations have found non mean-field exponents~\cite{Sharma2016,sharma2016strain,vermeulen2017geometry,shivers2019scaling,shivers2020nonlinear,arzash2020finite,arzash2021shear,arzash2022mechanics,lee2022stiffening}. Because our theory is a mean-field theory, it gives mean-field exponents by construction and cannot conclude the exponents of real networks. We leave the prediction of non-meanfield exponents for future work~\cite{Field} and focus here on getting the qualitative features of the transition. Below we show how these mean-field exponents are derived in our theory. 

For $\gamma>\gamma_c$, the network is stretch-dominated and $K_{_{\rm EM}}$ is independent of $\kappa$, allowing us to study the critical behavior for $\kappa=0$. In this case the nonlinear springs are described by the Landau-like energy of Eq.~(\ref{e304}). We first consider the network at the critical strain $\gamma=\gamma_c^+$, where all springs are stretched to their critical displacement $v_c$. For a strain slightly above the critical strain, $\gamma=\gamma_c+\Delta \gamma$, the springs are stretched to $v_c+\Delta v$. The value of $\Delta v$ may vary for each spring, while $\Delta v\sim\Delta \gamma$ holds for all springs. The differential spring constant, see Eq.~(\ref{e305}), is Taylor-expanded to linear order in $\Delta v$ as $k_{\rm AH}= k_c+k'_{\rm AH}(v_c)\Delta v$. Such a dependence leads to a similar relation for the macroscopic network elasticity with $K_{_{\rm EM}}=K_c+K_{_{\rm EM}}'\Delta \gamma$, where $K_{_{\rm EM}}'$ is a constant. Therefore, above the critical strain we have $K_{_{\rm EM}}-K_c\sim \mu |\Delta \gamma|^1$. 

For $\gamma<\gamma_c$, the network is bend-dominated, allowing us to consider the extreme limit $\mu\to\infty$, where no stretching deformation is allowed. In this case $U_{\rm AH}=0$ for $|\bm v_{\rm NA\perp}|\leq v_c$ and $U_{\rm AH}=\infty$  for $|\bm v_{\rm NA\perp}|>v_c$, such that $U_{\rm SP}$ is a harmonic potential for $|\bm v_{\rm NA\perp}|\leq v_c$ and $U_{\rm SP}=\infty$  for $|\bm v_{\rm NA\perp}|>v_c$. This suggests that each spring can only deform in the region $|\bm v_{\rm NA\perp}|\leq v_c$ and resists any displacement larger than $v_c$. In Appendix~\ref{Ab} we show that the spring energy leads to an unusual fiber conformation near the critical strain that results in $K_{_{\rm EM}}\sim \kappa|\Delta \gamma|^{-3/2}$. Interestingly, in Ref.~\cite{Rens2018} the same exponent is derived by analyzing the floppy deformation modes in the network.  

\begin{figure}[t]
	\centering
	\includegraphics[scale=0.43]{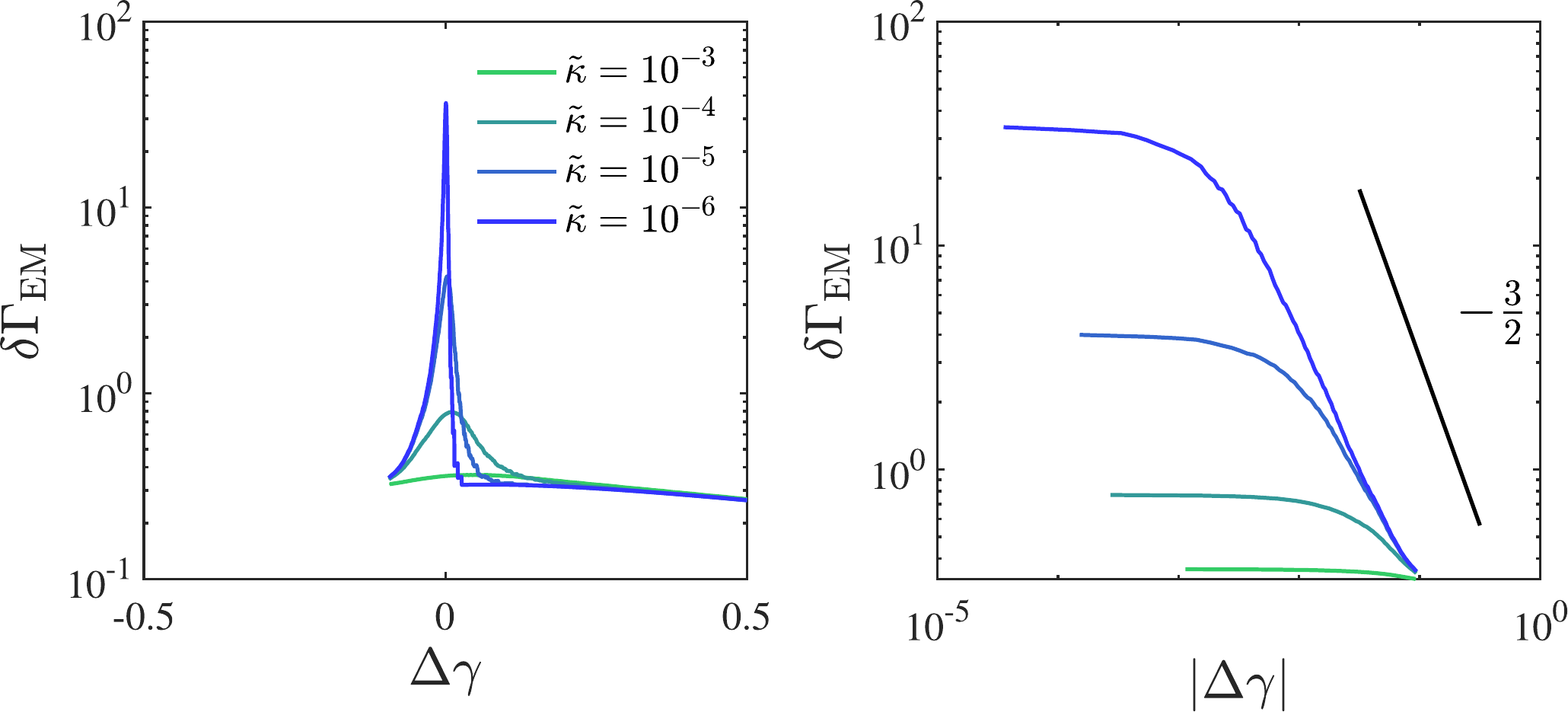}
	\caption{(a) Non-affine fluctuations $\delta \Gamma$ as function of $\Delta \gamma=\gamma-\gamma_c$, for various values of $\tilde \kappa = \kappa/(\mu\ell_c^2)$. The fluctuations diverge at the critical point. (b) Scaling behavior of  $\delta \Gamma$ for $\gamma<\gamma_c$. The network parameters are the same as in Fig.~\ref{K}. The color code used in (b) is the same as that in (a).  }
	\label{fluctuations}
\end{figure}
\subsection{Non-affine fluctuations}
Previous simulations and scaling theory have revealed another important quantity in characterizing the mechanical criticality, which is the non-affine fluctuations, or the differential non-affinity~\cite{Head2003,didonna2005nonaffine,Sharma2016,shivers2019scaling}. This quantity is defined in the original network as 
\begin{equation}
\begin{aligned}
\delta \Gamma_{_O} = \frac{1}{\ell_c^2\delta \gamma^2}\langle |\delta {\bm u}_{\rm NA}|^2\rangle\,.
\end{aligned}
\label{e16}
\end{equation} 
Here $\delta {\bm u}_{\rm NA}$ is the non-affine displacement of a single node in the original network due to a small incremental strain $\delta \gamma$, and the average is with respect to all nodes in the network. To estimate $\delta \Gamma_{_O}$, we similarly define the non-affine fluctuations of the EMT network:
\begin{equation}
\begin{aligned}
\delta \Gamma_{_{\rm EM}} = \frac{1}{\ell_c^2\delta \gamma^2}\langle |\delta {\bm v}_{\rm NA}|^2\rangle\,.
\end{aligned}
\label{e17}
\end{equation} 

In Fig.~\ref{fluctuations} (a) we plot $\delta \Gamma_{_{\rm EM}}$ as function of $\Delta \gamma = \gamma-\gamma_c$. For all $\tilde \kappa$ values, we find non-monotonic dependencies of $\delta \Gamma_{_{\rm EM}}$ on $\Delta \gamma$  with a sharp peak near $\Delta \gamma=0$. The height of the peak increases for decreasing $\tilde \kappa$ and diverges for $\tilde \kappa\to0$ when $\Delta \gamma<0$. Such divergence in fluctuations is an important signal of a critical point, which has been observed in previous simulations~\cite{sharma2016strain,shivers2019scaling}. For $\Delta \gamma<0$, we observe $|\delta \Gamma_{_{\rm EM}}|\sim|\Delta \gamma|^{-\lambda}$ with an exponent equal to $\lambda=\phi-f=1.5$, agreeing with previous simulations and scaling theory~\cite{BroederszThesis,Rens2018,shivers2019scaling}.  

For $\Delta \gamma>0$, however, we do not observe a divergence of the non-affinity, which appears to contradict both prior simulations and the general expectation of equal exponents above and below a critical point~\cite{shivers2019scaling}.  Nevertheless, such an apparent contradiction is expected in our framework, because our theory does not capture accurately the non-affinity above the critical point. By construction, our theory guarantees that the EMT network approximates the elasticity of the original network, i.e., $K_{_{\rm EM}}/ K_{O}\simeq 1$. On the other hand, $\delta \Gamma$ can be interpreted as a susceptibility-like quantity, $\delta \Gamma\sim K^\kappa(0,\gamma)$, where $K^{\kappa}(\kappa,\gamma)\equiv\partial K(\kappa,\gamma)/\partial \kappa$~\cite{shivers2019scaling}.  $\delta \Gamma$ thus reflects the $\kappa$-dependence of the elasticity. In the small $\kappa$ limit, we have
\begin{equation}
\begin{aligned}
K_{_{\rm EM}}(\kappa,\gamma)&=K_{_{\rm EM}}(0,\gamma)+\kappa K^\kappa_{_{\rm EM}}(0,\gamma) ,\\
K_{O}(\kappa,\gamma)&=K_{O}(0,\gamma)+\kappa K^\kappa_{_O}(0,\gamma). 
\end{aligned}
\label{e18}
\end{equation} 
Below the critical point, because $K(0,\gamma)=0$, the condition $K_{_{\rm EM}}/ K_{_O}\approx 1$ naturally leads to $K^\kappa_{\rm EM}(0,\gamma)/K^\kappa_{_O}(0,\gamma)\approx1$, hence $\delta \Gamma_{_{\rm EM}}/\delta \Gamma_{_O}\approx1$. Together with $K_{_{\rm EM}}\approx K_{_O}$ we have $\delta \Gamma_{_{\rm EM}}\approx \delta \Gamma_{_O}$. However, above the critical point we have $K(0,\gamma)\sim \mu$, which represents the dominating part in $K$ that is independent of $\kappa$. In this case, the condition $K_{_{\rm EM}}/ K_{_O}\approx 1$ only guarantees that $K_{_{\rm EM}}(0,\gamma)/K_{_O}(0,\gamma)\approx1$ and no longer guarantees similar $\kappa$-dependencies of $K_{_{\rm EM}}$ and $K_{_O}$. Therefore,  the theoretical prediction of $\delta \Gamma_{_{\rm EM}}$ becomes inaccurate for $\gamma>\gamma_c$. 

\section{Discussion and Conclusion}
\label{sec5}
In this work we have presented a non-linear EMT that analytically captures the mechanical critical phase transition of fiber networks. For this, we have extended our previous linear EMT~\cite{Linear} for non-affine deformations by introducing a phenomenological anharmonic spring energy that exhibits a Landau-like structure.  Such modification of the spring energy is sufficient to capture the strain-controlled mechanical phase transition of fiber networks, with good overall agreement with previous numerical and experimental results. Our results show a discontinuous transition of the differential elasticity $K_{_{\rm EM}}$ in the non-bending limit ($\kappa=0$), which agrees with previous simulations~\cite{vermeulen2017geometry,merkel2019minimal,arzash2020finite}. We also predict critical exponents and diverging non-affine fluctuations in the vicinity of the critical point, within the limits expected of a mean-field theory. 

Our previous linear EMT is based on two key linear relations: the first one is the linear force-extension relation of the harmonic springs (e.g., in Eq.~(\ref{e4})), the second one is a linear relation between the microscopic and macroscopic deformation (e.g., in Eq.~(\ref{e7}))~\cite{Linear}. In calculation of the nonlinear elasticity, both of these two relations may need nonlinear corrections. However, in our nonlinear EMT we have only taken into account a nonlinear correction to the spring energy, while the linear microscopic-macroscopic deformation relation remains unchanged. The reason for this assumption is that the microscopic-macroscopic deformation relation is a geometric property of the network. Therefore, one would expect it to be  dominated by the nonlinear terms only for $\gamma$ approaching unity. For fiber networks, both numerical and experimental studies have observed phase transitions at strain much smaller than one, suggesting that the nonlinear contribution in the deformation relation can be neglected. By contrast, the spring energy reflects a local mechanical property of the network, whose nonlinear contribution could dominate even at small strain~\cite{Broedersz2014}. In fact, we predict the critical strain due to the anharmonic spring energy to be $\gamma_c\sim v_c/L\sim \ell_c^2/L^2$, which is indeed small for long fibers. \SC{Both prior experiments and experimentally-motivated simulations have shown the critical strain values to be in the $10\%-30\%$ range~\cite{Sharma2016}, which is consistent with our small-strain assumption.}

Another nonlinearity of biopolymers lies in their longitudinal force-extension relation. For athermal fibers such nonlinearity is seen through  their buckling under compression. The buckling of fibers, however, is absent in the presented work, because we neglect the coupling between transverse and longitudinal displacements (see below Eq.~(\ref{e2})). As a result, our EMT fails to capture the  non-zero normal stresses in real networks that is induced by fiber buckling~\cite{janmey2007negative,heussinger2007nonaffine,conti2009cross,shivers2019normal}. In future work, it will be interesting to extend the EMT to study the nonlinear elasticity of  thermal semiflexible polymer networks induced by the  nonlinear and asymmetric force-extension relation of fibers~\cite{Thermal}. \SC{We also note that our EMT can be extended to other modes of deformation beyond simple shear. For a general deformation tensor $\uubar{\bm \Lambda}_{_{\rm EM}}$, one simply needs to modify the total energy above to $E_{_{\rm EM}}=H_{_{\rm EM}}-V\uubar{\bm \Lambda}_{_{\rm EM}}:\uubar{\bm \Sigma}_{_{\rm EM}}$, with $\uubar{\bm \Sigma}_{_{\rm EM}}$ being the stress tensor. }

The traditional, lattice-based EMT has been extensively applied in studying the elasticity of 2D networks. Despite the great success in predicting both linear and even some nonlinear~\cite{Sheinman2012} properties, a traditional EMT faces limitations when describing strain-controlled criticality. This is because the criticality is intrinsically related to non-affine deformations, which are absent in traditional EMT approaches that are based on a perfect lattice as the EMT network~\cite{phillips1979topology,thorpe1983continuous,Feng1985,Das2007,broedersz2011criticality,Sheinman2012,Mao2013,Mao20132}. The present work reproduces non-affine deformations or fluctuations of networks and even quantitative aspects of their divergence near the critical point, although our model appears to underestimate the non-affinity, especially in the stretch-dominated regime (see Fig.~\ref{fluctuations}). The fact that non-affine fluctuations are accounted for in a homogeneous EMT network may seem counter-intuitive. The reason it works is that the basic element in the EMT is not a dimensionless particle, but a fiber with finite length and a specific orientation. Thus, the microscopic deformation of each node can still vary according to the orientation ${\hat{\bm n}}$ and the position along the fiber $s$. It is this variation that gives rise to the non-affine fluctuations in our EMT. Moreover, even homogeneous networks can exhibit non-affine deformations under shear: One example of this is the perfect central-force honeycomb lattice that is subisostatic and has vanishing linear elasticity. 

The bending stiffness $\kappa$ can be regarded as a {\it stabilization factor}: When $\kappa=0$, fiber networks reduce to spring networks that are floppy to linear deformation. The floppy spring networks are thus stabilized by finite $\kappa$ values in linear regime. Other physical quantities may also serve as  { stabilization factors} for spring networks. For example, a finite temperature also leads to an entropic linear elasticity of spring networks~\cite{plischke1998entropic,dennison2013fluctuation,mao2015mechanical}. Our EMT may be extended to study the mechanical phase transition of spring networks in the presence of such other {stabilization factors} as well. The spring energy $U_{\rm SP}$ may need to be modified according to the specific {stabilization factor}, but its anharmonic part should remain unchanged, because it is controlled by the stretch rigidity alone. 

The nonlinear EMT presented in this paper is a first step towards a theoretical understanding of the nature of the critical phase transition in fiber networks. While the mean-field behavior of the phase transition is predicted by the EMT, it would be interesting to consider a field theory which goes beyond mean-field~\cite{Field}, as non mean-field exponents have been reported repeatedly in numerical simulations~\cite{Sharma2016,sharma2016strain,vermeulen2017geometry,shivers2019scaling,shivers2020nonlinear,arzash2020finite,arzash2021shear,arzash2022mechanics,lee2022stiffening}.

\section*{acknowledgements}
This work was supported in part by the National
Science Foundation Division of Materials Research
(Grant No. DMR-2224030) and the National Science
Foundation Center for Theoretical Biological Physics
(Grant No. PHY-2019745).

\appendix
\counterwithin{figure}{section}
\counterwithin{equation}{section}

\section{Derivation of anharmonic spring energy}
\label{A1}

In this appendix we derive the anharmonic spring energy $U_{\rm AH}$ in the infinite molecular weight limit (Eq.~(\ref{e301})). 

The spring energy is determined by Eq.~(\ref{e5a}), which is essentially a coherent potential approximation (CPA). As we show in the supplementary material of Ref.~\cite{Linear}, the CPA is equivalent to the test-force in which one exerts
\begin{figure}[b]
	\centering
	\includegraphics[scale=0.3]{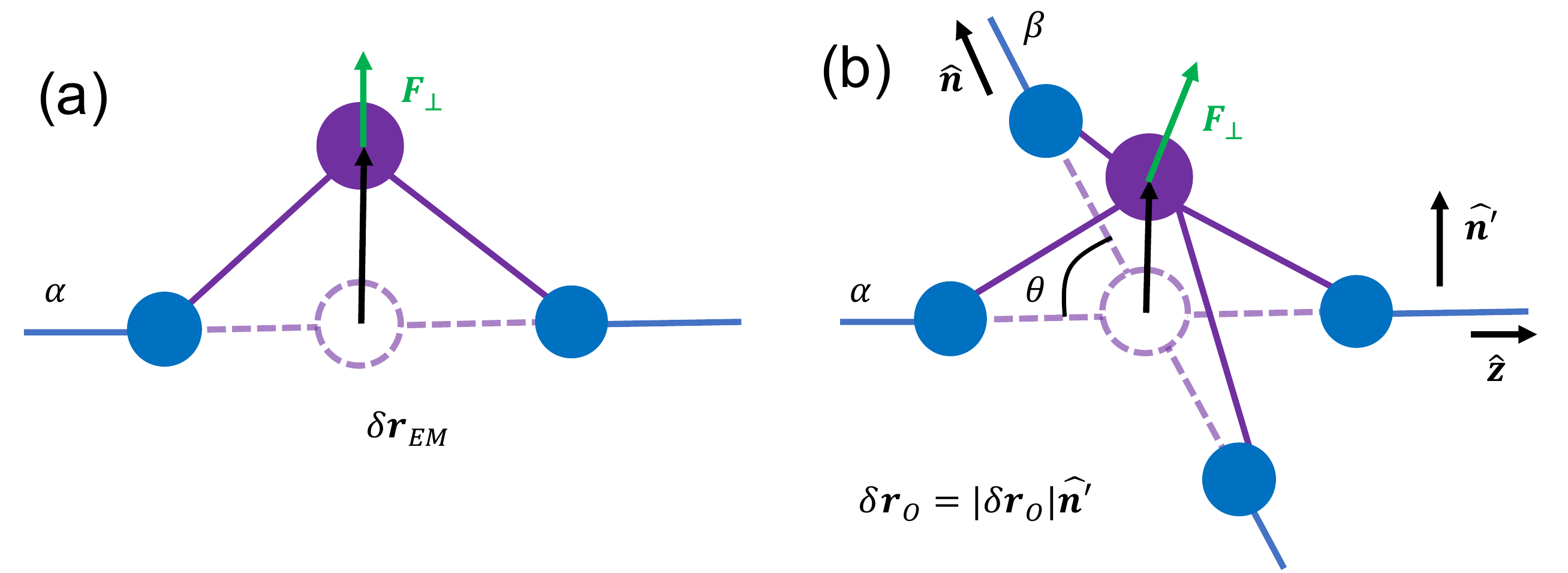}
	\caption{Sketch of the test force approach. A particular node on the purple fiber is deformed by a test force $\bm F$. The resulting displacement is $\delta \bm r_{\rm EM}$ in the EMT  (a), and $\delta \bm r_O$ in the original network (b). The adjacent nodes are assumed to be fixed for the infinite molecular weight limit, which is not the case for finite molecular weight. 
	}
	\label{Fig.A1}
\end{figure}
a test force at a particular crosslink on the $\alpha$-th fiber in both the original network and the EMT, and calculate the resulting displacements $\delta {\bm r}_O$ and $\delta {\bm r}_{\rm EM}$. Because we are only interested in the transverse response of the spring, we set the direction of the force to be perpendicular to the $\alpha$-th fiber and denote the force by ${\bm F}_\perp$.   In the original network because the crosslink is also connected to another fiber (the $\beta$-th), $\delta {\bm r}_O$ would also depend on the orientation of the other fiber $\hat{\bm n}$. We use $\langle \delta {\bm r}_O\rangle_{\hat{\bm n}}$ to denote the average of $\delta {\bm r}_O$ with respect to the distribution of  $\hat{\bm n}$. To ensure that the EMT gives the same average response as the original network, we impose
\begin{equation}
\begin{aligned}
\delta {\bm r}_{\rm EM}=\langle \delta {\bm r}_O\rangle_{\hat{\bm n}}\,.
\end{aligned}
\label{eA01}
\end{equation} 
Note that in our previous linear EMT Eq.~(\ref{eA01}) is only analyzed in the small force limit. In this nonlinear EMT Eq.~(\ref{eA01}) should hold for arbitrary force. In the calculation below we also assume the adjacent nodes (blue ones in Fig.~\ref{Fig.A1}) do not move, which corresponds to the high molecular weight limit, as explained in the main text. 

Let us start with the EMT and write the stretching energy $H_{s}\left[{\bm u}(s)\right]=\mu \int \diff s (|\hat {\bm n}+\partial {\bm u}/\partial s|-1)^2/2$ in the coarse grained limit, i.e., approximating the fiber with straight segments connected by crosslinks:
\begin{equation}
\begin{aligned}
H_s = \frac{\mu}{2\ell_c}\sum_i (\sqrt{|\ell_c\hat{\bm n} +\bm u_i-\bm u_{i-1}|^2}-\ell_c)^2\,,
\end{aligned}
\label{eA09}
\end{equation} 
where $\bm u_i$ is the displacement of the $i$-th crosslink. In the test-force approach only a particular crosslink (the $j$-th) is deformed, corresponding to  $\bm u_j=\delta{\bm r}_{\rm EM}$ and $\bm u_i=0$ for other crosslinks. 

The leading order term of $H_s$ in $\delta{\bm r}_{\rm EM}$ is
\begin{equation}
\label{eA19}
H_s=\left\{
\begin{aligned}
&\frac{\mu}{\ell_c}(\delta {\bm r}_{\rm EM}\cdot \hat{\bm n})^2   &\delta {\bm r}_{\rm EM}\cdot \hat{\bm n}\neq 0  \\
&\frac{\mu}{4\ell_c^3}|\delta {\bm r}_{\rm EM}|^4   & \delta {\bm r}_{\rm EM}\cdot \hat{\bm n}= 0,
\end{aligned}
\right.
\end{equation}
which is of fourth order if $\delta{\bm r}_{\rm EM}$ is perpendicular to the fiber and of second order otherwise. Because  the force ${\bm F}_{\perp}$ is perpendicular to $\hat{\bm n}$, the leading order term in Eq.~(\ref{eA09}) is $H_s=\frac{\mu}{4\ell_c^3}|\delta {\bm r}_{\rm EM}|^4$. 
Similarly we write the bending energy $H_{b}\left[{\bm u}(s)\right]=\kappa \int \diff s |\partial ^2{\bm u}_\perp/\partial s^2|^2/2$ in the coarse grained limit as:
\begin{equation}
\label{eA21}
H_b=
\frac{3\kappa}{\ell_c^3}\delta {\bm r}_{\rm EM}^2. 
\end{equation}

The total energy associated with the displacement is the sum of the bending, stretching and the spring energy:
\begin{equation}
\begin{aligned}
\Delta E_{_{\rm EM}}=\frac{12\kappa}{\ell_c^3}|\delta {\bm r}_{\rm EM}|^2+ U_{\rm AH}(\delta {\bm r}_{\rm EM})+\frac{\mu}{4\ell_c^3}|\delta {\bm r}_{\rm EM}|^4-{\bm F}_\perp \cdot \delta {\bm r}_{\rm EM},
\end{aligned}
\label{eA02}
\end{equation} 
where the first term is the bending energy, the second term is the anharmonic spring energy, the third term is the stretching energy of the two purple bonds (keeping the leading term) and the fourth term is the work done by the force. The resulting $\delta {\bm r}_{\rm EM}$ of the force is found by minimizing $\Delta E_{_{\rm EM}}$. Because of the rotational symmetry $U_{\rm AH}$ should be a function of $|\delta {\bm r}_{\rm EM}|^2$, which further leads to a Taylor expansion $U_{\rm AH}=a_2|\delta {\bm r}_{\rm EM}|^2+a_4|\delta {\bm r}_{\rm EM}|^4+...$. Since $U_{\rm AH}$ does not contribute to the linear elasticity of the spring, we have $a_2=0$ and the leading term in $U_{\rm AH}$ is the quartic term. In this case $U_{\rm AH}=a_4|\delta {\bm r}_{\rm EM}|^4$ is determined by $a_4$, and in the minimum energy state we have
\begin{equation}
\begin{aligned}
\frac{24\kappa}{\ell_c^3}|\delta {\bm r}_{\rm EM}|+\left(4a_4+\frac{\mu}{\ell_c^3}\right)|\delta {\bm r}_{\rm EM}|^3-|{\bm F}_\perp|=0\,.
\end{aligned}
\label{eA03}
\end{equation} 
While the exact solution of Eq.~(\ref{eA03}) is cumbersome, one can analyze the solution under extreme limits. For $|{\bm F}_\perp|\ll \kappa^{3/2}\ell_c^{-3}(4a_4\ell_c^3+\mu)^{-1/2}$, the linear term of $|\delta {\bm r}_{\rm EM}|$ dominates and $|\delta {\bm r}_{\rm EM}|=|{\bm F}_\perp|\ell_c^3/(24\kappa)$ is independent of $a_4$. For $|{\bm F}_\perp|\gg \kappa^{3/2}\ell_c^{-3}(4a_4\ell_c^3+\mu)^{-1/2}$, the cubic term of $|\delta {\bm r}_{\rm EM}|$ dominates and 
\begin{equation}
\begin{aligned}
|\delta {\bm r}_{\rm EM}|=|{\bm F}_\perp|^{1/3}(4a_4\ell_c^3+\mu)^{-1/3}\ell_c. 
\end{aligned}
\label{eA22}
\end{equation} 
Therefore, in order to determine $U_{\rm AH}$ one should focus on the large $|{\bm F}_\perp|$ limit.

We now switch to the original network. Without loss of generality, let the orientation of fiber $\alpha$ be $\hat {\bm z}=(0,0,1)$, ${\bm F}_\perp$ be ${\bm F}_\perp=F\hat{\bm x}$ where $\hat{\bm x} = (1,0,0)$, the orientation of fiber $\beta$ be $\hat{\bm n}=(\sin(\theta)\cos(\phi),\sin(\theta)\sin(\phi),\cos(\theta))$, and the displacement of the crosslink be $\delta {\bm r}_{O}$. According to Eq.~(\ref{eA19}), only when $\delta {\bm r}_{O}$ is perpendicular to both $\hat {\bm z}$ and $\hat {\bm n}$ is the leading order term in the stretching energy quartic. Otherwise the leading order term would be quadratic, which is larger than the quartic term as long as $|\delta {\bm r}_O|\ll\ell_c$. Therefore, in order to minimize the total energy, the displacement of the crosslink is restricted to the direction $\hat{\bm n}'=(\sin(\phi),-\cos(\phi),0)$, which is perpendicular to both $\hat{\bm z}$ and $\hat{\bm n}$. Letting $\delta {\bm r}_{O}=\delta r\hat{\bm n}'$, we write down the total energy:
 \begin{equation}
 \begin{aligned}
 \Delta E_O=\frac{6\kappa}{\ell_c^3}\delta {r}^2+\frac{\mu}{2\ell_c^3}\delta r^4-F\delta r \sin(\phi),
 \end{aligned}
 \label{eA04}
 \end{equation} 
Minimizing Eq.~(\ref{eA04}) leads to 
 \begin{equation}
\begin{aligned}
\frac{12\kappa}{\ell_c^3}\delta {r}+\frac{2\mu}{\ell_c^3}\delta r^3=F \sin(\phi)\,.
\end{aligned}
\label{eA20}
\end{equation}
Again, we analyze the solution of Eq.~(\ref{eA20}) in the large $F$ limit, which is
 \begin{equation}
\begin{aligned}
\delta {\bm r}_O=\left(\frac{F\sin(\phi)}{2\mu}\right)^{ 1/3}\ell_c\,\hat{\bm n}'\,.
\end{aligned}
\label{eA05}
\end{equation}
Taking the average of Eq.~(\ref{eA05}) with respect to the distribution of $\phi$, $P(\phi)=1/(2\pi)$ (due to the rotational symmetry) gives
 \begin{equation}
\begin{aligned}
\langle\delta {\bm r}_{O}\rangle_{\hat {\bm n}}=0.46\left(\frac{F}{\mu}\right)^{ 1/3}\ell_c\,\hat{\bm x}\,.
\end{aligned}
\label{eA06}
\end{equation}
Substituting Eq.~(\ref{eA06}) in Eq.~(\ref{eA01}) and using Eq.~(\ref{eA03}) leads to 
\begin{align}
U_{\rm AH}(\bm v_{\rm NA\perp})=2.32\frac{\mu}{\ell_c^3}|\bm v_{\rm NA\perp}|^4, 
\label{eA07}
\end{align}
which is Eq.~(\ref{e301}) of the main text.

\SC{
\section{Details of the energy minimization}
\label{Ac}
We describe here the details of the energy minimization of the EMT, i.e., minimizing the total energy in Eq.~(\ref{e9}),  
\begin{equation}
\begin{aligned}
E_{_{\rm EM}} = H_{_{\rm EM}}-\gamma_{_{\rm EM}} \sigma_{_{\rm EM}}\,.
\end{aligned}
\label{eC01}
\end{equation} 
In principle, one needs to minimize Eq.~(\ref{eC01}) with the constraints that all other components of the deformation tensor being zero. However, as we show below, these constraints are naturally satisfied in our work, hence we are allowed to minimize Eq.~(\ref{eC01}) without constraints. 
\\
Because our goal is to minimize $E_{_{\rm EM}}$ with respect to $\bm v^{\alpha}$, we start by writing $\gamma_{_{\rm EM}}$ in terms of $\bm v^{\alpha}$
\begin{align}
\gamma_{_{\rm EM}} &= \sum_\alpha \int \diff s{\bm t}_\perp^\alpha\cdot {\bm v}_{\rm NA\perp}^\alpha+\sum_\alpha \int \diff s{\bm t}_\parallel^\alpha\cdot {\bm v}_{\parallel}^\alpha\,,
\label{eC02}
\end{align}
where ${\bm t}_\perp^\alpha=5f_\perp(n_z^\alpha\hat{\bm x}-n_x^\alpha n_z^\alpha \hat{\bm n}^\alpha)$ and ${\bm t}_\parallel^\alpha=5f_\parallel n_x^\alpha n_z^\alpha \hat{\bm n}^\alpha$. Equation~(\ref{eC02}) is  derived from Eq.~(\ref{e7}), see Ref.~\cite{Linear}. As we discussed below Eq.~(\ref{e2}), it is convenient to introduce new variables $\partial \epsilon^\alpha/\partial s=\hat{\bm n}\cdot(\partial {\bm v}^\alpha_\parallel/\partial s)+|\partial {\bm v}^\alpha_\perp/\partial s|^2/2$. Here $\epsilon^\alpha$ describes the local stretch of the fiber. For small strain, we have ${\bm t}_\parallel^\alpha\cdot {\bm v}_{\parallel}^\alpha\approx {t}_\parallel^\alpha\epsilon^\alpha$, where ${t}_\parallel=5f_\parallel n_x^\alpha n_z^\alpha$, hence Eq.~(\ref{e7}) is written as
\begin{align}
\gamma_{_{\rm EM}} &= \sum_\alpha \int \diff s{\bm t}_\perp^\alpha\cdot {\bm v}_{\rm NA\perp}^\alpha+\sum_\alpha \int \diff s{\ t}_\parallel^\alpha\epsilon^\alpha\,,
\label{eC03}
\end{align}
It is instructive to write $H_{_{\rm EM}}$ in Eq.~(\ref{e2}) in terms of ${\bm v}_{\rm NA\perp}^\alpha$ and $\epsilon^\alpha$,
\begin{align}
H_{_{\rm EM}}&=\sum_{\alpha=1}^N\Big(H_{b}\left[{\bm v}_\perp^{\alpha}(s)\right]+H_{s}\left[{\bm v}^{\alpha}(s)\right]+H_K\left[\bm v_{\rm NA}^{\alpha}(s)\right]\Big)\notag\\&=\sum_{\alpha=1}^N\Big(H_{b}\left[{\bm v}_{\rm NA\perp}^{\alpha}(s)\right]+H_{s}\left[\epsilon^{\alpha}(s)\right]+H_K\left[\bm v_{\rm NA \perp}^{\alpha}(s)\right]\Big)\,.
\label{eC04}
\end{align}
In the second equality we use the fact that in affine deformations no bending energy evolves, such that $H_{b}\left[{\bm v}_\perp^{\alpha}(s)\right]=H_{b}\left[{\bm v}_{\rm NA\perp}^{\alpha}(s)\right]$. 
Substituting Eqs.~(\ref{eC03}) and (\ref{eC04}) into Eq.~(\ref{eC01}), we find that ${\bm v}_{\rm NA\perp}^\alpha$ and $\epsilon^\alpha$ are decoupled in $E_{_{\rm EM}}$. The minimization is then performed through $\delta E_{_{\rm EM}}/\delta {\bm v}_{\rm NA\perp}^\alpha=0$ and $\delta E_{_{\rm EM}}/\delta \epsilon^\alpha=0$, which leads to
\begin{subequations}
\begin{align}
&\kappa\frac{\diff^4{\bm v}_{\rm NA\perp}^{\alpha}}{\diff s^4}+ \bm F_{K}({\bm v}_{\rm NA\perp}^\alpha) = \sigma_{_{\rm EM}} {\bm t}_\perp^{\alpha}\,,\label{eC05.1}\\
&\mu\frac{\diff^2{\epsilon}^{\alpha}_{\parallel}}{\diff s^2}=\sigma_{_{\rm EM}} { t}^{\alpha}_\parallel\,,\label{eC05.2}
\end{align}
\label{eC05}
\end{subequations} 
with natural boundary conditions (the boundary points are free to move, {i.e.,}  can take any value at the boundaries). 
In Eq.~(\ref{eC05.1}), the first term is the bending force, and the second term is the spring force $\bm F_{K}=\delta H_K\left({\bm v}_{\rm NA\perp}^\alpha\right)/\delta {\bm v}_{\rm NA\perp}^\alpha$. Surprisingly, we find that the bending force can be ignored: For small stress, because the spring force $\bm F_{K}$ is linear in ${\bm v}_{\rm NA\perp}^\alpha$, the solution of Eq.~(\ref{eC05.1}) is ${\bm v}_{\rm NA\perp}^\alpha(s)\sim s$, and there is no bending force; For large stress, the spring stiffens and $\bm F_{K}\sim \mu$ dominates, such that the contribution from the bending force can be neglected. Therefore, Eq.~(\ref{eC05}) is further simplified to 
\begin{subequations}
	\begin{align}
	& \bm F_{K}({\bm v}_{\rm NA\perp}^\alpha) = \sigma_{_{\rm EM}} {\bm t}_\perp^{\alpha}\,,\label{eC06.1}\\
	&\epsilon^{\alpha}_{\parallel}(s)=\frac{15\sigma_{_{\rm EM}}n_x^{\alpha}n_z^{\alpha}s}{NL \mu } \,.  \label{eC06.2}
	\end{align}
	\label{eC06}
\end{subequations}
Equation~(\ref{eC06.2}) is derived using the value of $f_\parallel$ in Eq.~(\ref{S210}). Due to the nonlinear nature of $\bm F_K$, Eq.~(\ref{eC06.1}) needs to be solved numerically. For each given $\sigma_{_{\rm EM}}$, the solution of Eq.~(\ref{eC06}) is then substituted into Eq.~(\ref{eC03}) to find the corresponding $\gamma_{_{\rm EM}}$. For large number of fibers $N$, the summation of all fibers in Eq.~(\ref{eC03})  is replaced by an integral with respect to an isotropic distribution of fiber orientation. Under spherical coordinates $\hat{\bm n}=(\sin(\theta)\cos(\phi),\sin(\theta)\sin(\phi),\cos(\theta) )$, the distributions of $\theta$ and $\phi$ are $P(\theta)=\sin(\theta)/2$ and $P(\phi)=1/(2\pi)$. 
\\
Having described the energy minimization process, let us check whether the solution satisfies the constraints that all non-$xz$ components of the deformation tensor are zero. First, the $yz$ and $xy$ components of the deformation tensor are zero because of the mirror symmetry in $y$ direction.  To show that the normal components ($xx$, $yy$ and $zz$) are also zero, we exerts two shear stresses $\sigma_{_{\rm EM}}$ and $-\sigma_{_{\rm EM}}$ on the network separately. By symmetry the normal components of the deformation tensor should be the same for these two opposite stresses. According to Eq.~(\ref{eC06}), we have $\bm v^\alpha(\sigma_{_{\rm EM}})=-\bm v^\alpha(-\sigma_{_{\rm EM}})$, hence, from Eq.~(\ref{e7}) $\uubar{\bm \Lambda}_{_{\rm EM}}(\sigma_{_{\rm EM}})=-\uubar{\bm \Lambda}_{_{\rm EM}}(-\sigma_{_{\rm EM}})$. Therefore, the normal components must be zero, and $\uubar{\bm \Lambda}_{_{\rm EM}}(\sigma_{_{\rm EM}})$ is a simple shear deformation, consistent with the constraints. 
\\
These constraints are naturally satisfied because we neglect fiber buckling by assuming a quadratic stretching energy, see below Eq.~(\ref{e2}). If fiber buckling is taken into account, these constraints must be explicitly taken into account in the minimization. In such a case, one may use the Lagrange multipliers method to find the non-zero normal stresses corresponding to the constraints.  It is known that there can be non-zero normal stress associated with the nonlinear stiffening for real networks~\cite{janmey2007negative,heussinger2007nonaffine,conti2009cross,shivers2019normal}. 
}

\SC{
	\section{$K_{_{\rm EM}}$ at the critical point}
	\label{Ad}
	In this section we plot the EMT results of the elasticity exactly at the critical point, $K_{_{\rm EM}}(\gamma_c)$, for various values of  $\tilde \kappa=\kappa/(\mu\ell_c^2)$, see Fig.~\ref{Fig.A3}. We find that $K_{_{\rm EM}}(\gamma_c)$ for $\kappa\to0$ is significantly smaller than its value for $\kappa=0$, implying that the small-$\kappa$ limit is qualitatively different from the zero-$\kappa$ case, consistent with the numerical evidence in Ref.~\cite{lerner2022scaling}.
	\begin{figure}[h]
		\centering
		\includegraphics[scale=0.45]{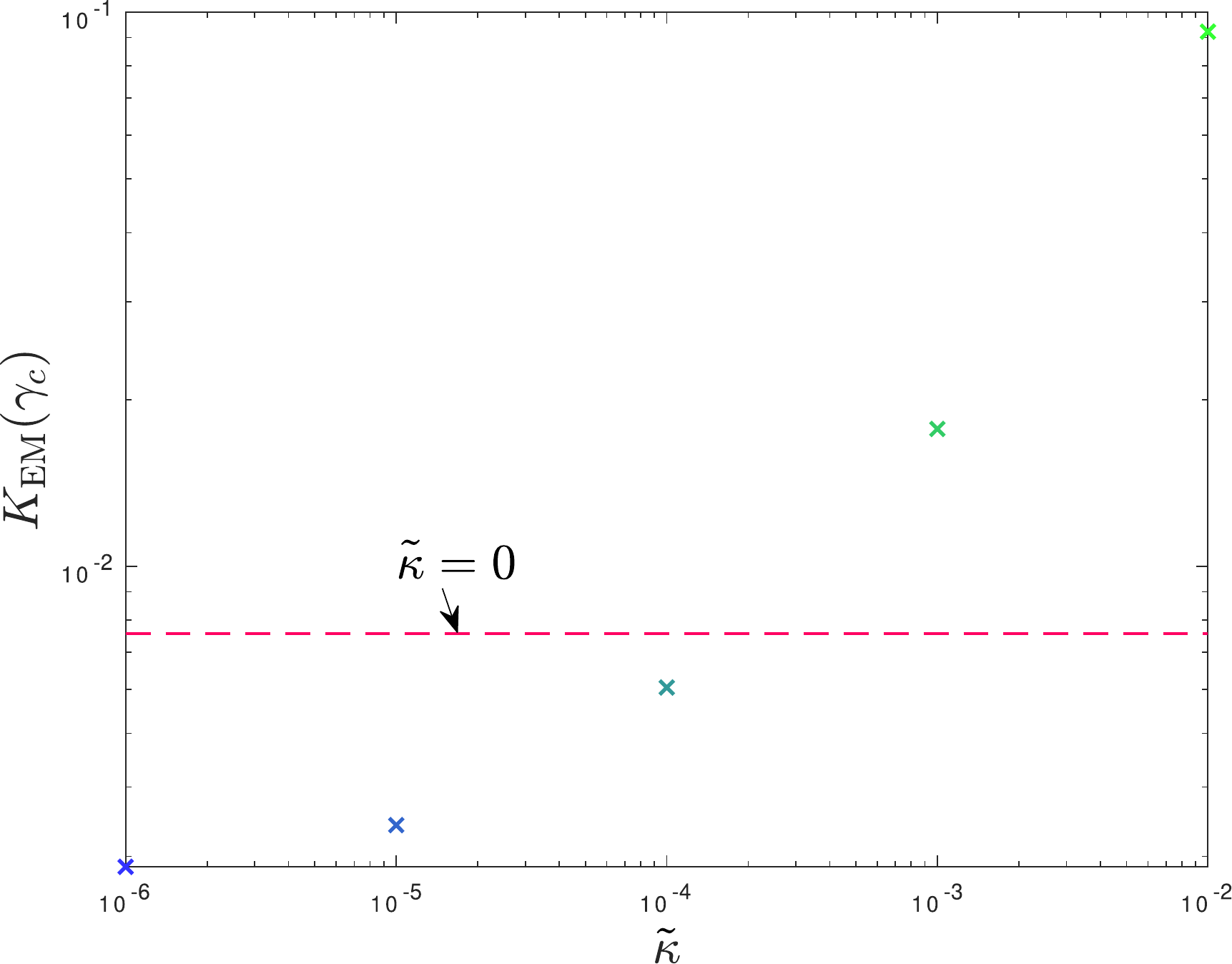}
		\caption{\SC{The elasticity of the EMT at the critical point, $K_{_{\rm EM}}(\gamma_c)$, as function of $\tilde \kappa$. The dashed line indicates the value of $K_{_{\rm EM}}(\gamma_c)$ for $\tilde \kappa=0$.  }}
		\label{Fig.A3}
	\end{figure}
}

\section{Derivation of the critical exponent $f-\phi$}
\label{Ab}
In this section we derive the critical exponent $f-\phi$, which determines the scaling behavior of $K$ when $\gamma<\gamma_c$. Such a regime is governed by the bending energy only, hence, the stretching energy and the longitudinal deformation of each fiber can be neglected, i.e., $\mu\to\infty$ and $\epsilon^\alpha=0$. Therefore, the microscopic-macroscopic deformation relation (Eq.~(\ref{eC03})) is also simplified to
\begin{align}
\gamma_{_{\rm EM}} &\simeq\sum_\alpha \int \diff s{\bm t}_\perp^\alpha\cdot {\bm v}_{\rm NA\perp}^\alpha \equiv\sum_\alpha\gamma^\alpha,
\label{eA08}
\end{align}
where ${\bm t}_\perp^\alpha=5f_\perp(n_z^\alpha\hat{\bm x}-n_x^\alpha n_z^\alpha \hat{\bm n}^\alpha)$. Here $\gamma_{_{\rm EM}}$ is decomposed to the strains of $N$ `one-fiber' systems, each with strain
\begin{align}
\gamma^\alpha &=  \frac{180 n_z^\alpha\hat{\bm x}-n_x^\alpha n_z^\alpha \hat{\bm n}^\alpha }{NL^3}\,\int \diff s\, s \bm v^\alpha_{\rm NA\perp}(s). 
\label{eA13}
\end{align}
In Eq.~(\ref{eA13}) we have used Eq.~(\ref{S210}) as the value of $f_\perp$. 

For simplicity we consider the $L\gg \ell_c$ limit, which allows us to write the spring energy $H_K$ in the continuum limit $H_K\left[\bm v^\alpha_{\rm NA\perp}\right]=(1/2)\int \diff s\, g(|\bm v^\alpha_{\rm NA\perp}(s)|)|\bm v^\alpha_{\rm NA\perp}(s)|^2$, where

\begin{figure}[t]
	\centering
	\includegraphics[scale=0.3]{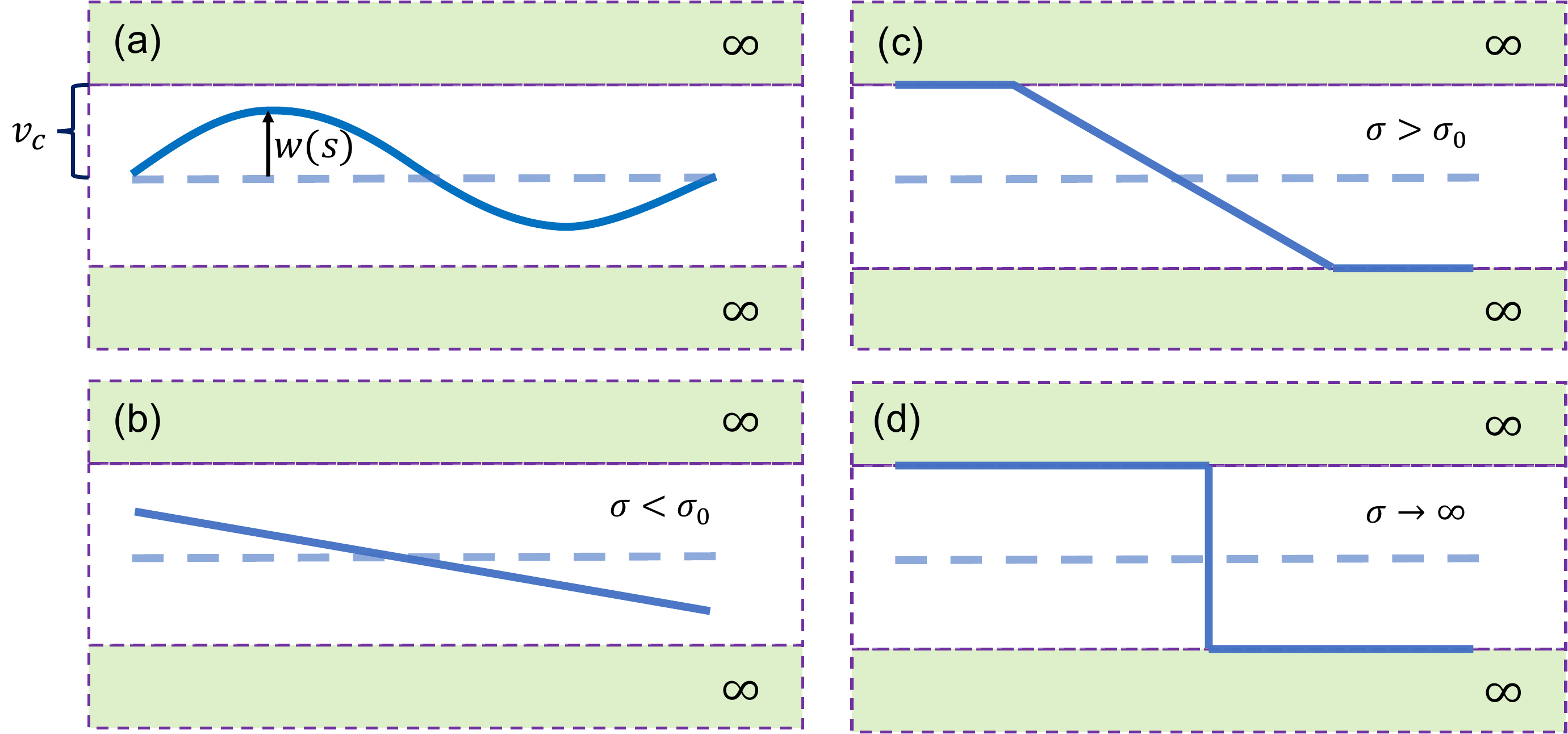}
	\caption{Illustration of the `one-fiber' system. (a) the system is formed by a single fiber with deformation field $w(s)$. The green region denotes where $H^*_K$ is infinity. (b) fiber deformation when $\sigma<\sigma_0$. (c) fiber deformation when $\sigma>\sigma_0$. (d) fiber deformation when $\sigma\to\infty$. 
	}
	\label{Fig.A2}
\end{figure}

\begin{equation}
\label{eA12}
g(|\bm v^\alpha_{\rm NA\perp}(s)|)=\left\{
\begin{aligned}
&\frac{18\kappa}{\ell_c^4}   & |\bm v^\alpha_{\rm NA\perp}(s)|\leq v_c \\
&\infty   & |\bm v^\alpha_{\rm NA\perp}(s)|>v_c
\end{aligned}
\right.
\end{equation}
describes a stiffness density of the effective springs. Note that the stiffness is approximated to be  infinity if $|\bm v^\alpha_{\rm NA\perp}(s)|>v_c$, because the in $H_K$ a term proportional to the stretching rigidity enters when $|\bm v^\alpha_{\rm NA\perp}(s)|>v_c$. In this case the minimization of the energy is found through Eq.~(\ref{eC06.1}), which leads to deformation behavior that is different for stress below and above a threshold value $\sigma_\alpha=36\kappa v_c/(bL\ell_c^4)$, where $b={180| n_z^\alpha\hat{\bm x}-n_x^\alpha n_z^\alpha \hat{\bm n}^\alpha| }/({NL^3})$. For $\sigma<\sigma_\alpha$, all parts of the fiber obey $|\bm v^\alpha_{\rm NA\perp}(s)|<v_c$, corresponding to the white region of Fig.~\ref{Fig.A2}. For $\sigma\geq \sigma_\alpha$, some parts of the fiber stay at the boundary between the white and green regions of Fig.~\ref{Fig.A2} ($|\bm v^\alpha_{\rm NA\perp}(s)|=v_c$). The resulting strain is
\begin{equation}
\label{eA14}
\gamma^\alpha=\left\{
\begin{aligned}
&A\sigma_{_{\rm EM}}   & \sigma_{_{\rm EM}}<\sigma_\alpha \\
&\gamma_c^\alpha-B/\sigma^2  & \sigma_{_{\rm EM}}\geq\sigma_\alpha,
\end{aligned}
\right.
\end{equation}
where $A=({b^2\ell_c^4})/({216\kappa L^3})$, $\gamma_c^\alpha=bv_c/(4L)$ and $B=({108\kappa^2v_c^3 L^3})/(b\ell_c^8)$. Here, $\gamma_c^\alpha$ is the maximum strain of the `one-fiber' system in the $\sigma\to\infty$ limit, which corresponds to a critical strain of the network, because it is the largest possible strain without causing stretching deformation. The corresponding nonlinear elasticity of the `one-fiber' system is ($K^\alpha=\diff \sigma/\diff \gamma^\alpha$)
\begin{equation}
\label{eA15}
K^\alpha=\left\{
\begin{aligned}
1/A & , & \sigma_{_{\rm EM}}<\sigma_\alpha \\
2\sigma_{_{\rm EM}}^3/B & , & \sigma_{_{\rm EM}}\geq\sigma_\alpha.
\end{aligned}
\right.
\end{equation}

Equation~(\ref{eA14}) suggests that for $\sigma<\sigma_\alpha$ the `one-fiber' system shows linear elasticity. When $\sigma>\sigma_\alpha$ the system stiffens non-linearly and we have $K^\alpha\sim(\gamma^\alpha_c-\gamma^\alpha)^{-3/2}$. Thus, we find that $f-\phi=-3/2$ in the `one-fiber' system. 

Having derived the scaling exponent for the `one-fiber' system, let us go back to the EMT network. According to Eq.~(\ref{eA15}), for each `one-fiber' system we have $\gamma_{c}^\alpha-\gamma^{\alpha}\sim \sigma^{-2}$ for $\sigma$ large enough. Therefore, for the EMT network we have 
\begin{align}
\gamma_c-\gamma_{_{\rm EM}}&=\sum_\alpha\left[\gamma_{c}^\alpha-\gamma^\alpha \right]\sim \sigma_{_{\rm EM}}^{-2}\,,
\label{eA18}
\end{align}
and hence $K_{_{\rm EM}}\sim(\gamma_c-\gamma_{_{\rm EM}})^{-3/2}$.

\bibliography{citation}
\bibliographystyle{rsc}

\end{document}